\documentclass[twocolumn,secnumarabic,amssymb, nobibnotes, aps, prd]{revtex4-2}

\usepackage[colorlinks=true,allcolors=blue]{hyperref}       
\usepackage{amsmath}
\usepackage{natbib}
\usepackage{adjustbox}
\usepackage{subfigure}
\usepackage{graphicx}
\usepackage{booktabs}       
\usepackage{amsfonts}       
\usepackage{nicefrac}       
\usepackage{microtype}
\usepackage{color}
\usepackage{textcomp}
\usepackage{gensymb}
\usepackage{cleveref}
\usepackage{multirow}
\usepackage{makecell}
\usepackage{array}
\newcolumntype{C}[1]{>{\centering\arraybackslash}m{#1}}

\begin{document}

\title[]{Loop Quantum Gravity motivated multihorizon rotating black holes}
\author{Jitendra Kumar $^{a}$ } \email{jitendra0158@gmail.com} 
\author{Shafqat Ul Islam$^{a}$ } \email{shafphy@gmail.com}

\author{Sushant~G.~Ghosh$^{a,\;b}$} \email{sghosh2@jmi.ac.in, sgghosh@gmail.com}
\affiliation{$^{a}$ Centre for Theoretical Physics, 
	Jamia Millia Islamia, New Delhi 110025, India}
\affiliation{$^{b}$ Astrophysics and Cosmology Research Unit, 
	School of Mathematics, Statistics and Computer Science, 
	University of KwaZulu-Natal, Private Bag 54001, Durban 4000, South Africa}

\date{\today}

\begin{abstract}
With a semiclassical polymerization in the loop quantum gravity (LQG), the interior of the Schwarzschild black holes provides a captivating single-horizon regular black hole spacetime. 
The shortage of rotating black hole models in loop quantum gravity (LQG) substantially restrains the progress of testing LQG from observations. Motivated by this, starting with a spherical LQG black hole as a seed metric, we construct a rotating spacetime using the revised Newman-Janis algorithm, namely, the LQG-motivated rotating black holes (LMRBH), which encompasses Kerr ($l=0$) black holes as an exceptional case. We discover that for any random  $l>0$, unlike Kerr black hole, an extremal LMRBH refers to a black hole with angular momentum $a>M$. The rotating metric, in parameter space,  describes (1) black holes with an event and Cauchy horizon, (2) black holes with three horizons, (3) black holes with only one horizon or  (4) no horizon spacetime.  We also discuss the horizon and global structure of the LMRBH spacetimes and its dependence on $l/M$ that exhibits rich spacetime structures in the ($M,\;a,\;l$) parameter space. 
\end{abstract}
\maketitle

\section{Introduction}\label{Sec-1}
That the gravitational collapse of an adequately massive star ($\sim3.5M_{\odot}$) inevitably guides to a spacetime singularity is a reality demonstrated by an elegant theorem due to Hawking and Penrose \cite{Hawking:1970zqf} (see also Hawking and Ellis \cite{Hawking:1973}). However, it is a general belief that these singularities do not live in nature but are an artefact of classical general relativity. By its very definition, the existence of a singularity conveys spacetime failing to exist, signalling a breakdown of the laws of physics. Thus, singularities must be replaced by other objects in a more unified theory for these laws to exist. The extreme condition at the singularity indicates that one should count on quantum gravity, developed to fix this singularity \cite{Wheeler:1964}. In particular, recent work indicates that loop quantum gravity (LQG) may be capable of resolving the singularities \cite{Ashtekar:2006wn,Ashtekar:2006es,Vandersloot:2006ws}. Because of the intrinsic problem in solving the whole system, the emphasis is on spherically symmetric black hole spacetimes \cite{Modesto:2004wm,Ashtekar:2005qt,Modesto:2005zm}.

LQG is a non-perturbative quantum gravity technique within which the discretized eigenvalues of a set of geometrical operators handle the core of spacetime. Remarkably, the non-zero area gap associated with the geometrical operators is crucial in removing classical singularities \cite{Bojowald:2020dkb}.  It turns out that within the framework of LQG and considering the semi-classical regimes, one can construct various effective but spherical symmetric black hole models such that singularity occuring in the general relativity is  now substituted by a transition surface that connects a black hole and a white hole region \cite{Gambini:2013ooa,Corichi:2015xia,Olmedo:2017lvt,Ashtekar:2018lag,Ashtekar:2018cay,Bodendorfer:2019xbp,Bodendorfer:2019cyv,Arruga:2019kyd,Assanioussi:2019twp,BenAchour:2020gon,Gambini:2020nsf,Bodendorfer:2019nvy,Bodendorfer:2019jay,Blanchette:2020kkk,Assanioussi:2020ezr,Chen:2022nix}. The most successful
endeavour is the phase space quantization, i.e., the  
polymerization techniques formulated in 
LQG, which settles the big-bang
singularity \cite{Bojowald:2005epg,Singh:2009mz,Zhang:2007yu}. It is naturalistic to represent the black-hole spacetimes by assuming the quantum corrections introduced by polymerization schemes. Especially the semi-classical polymerization technique, that preserves the LQG's idea of spacetime discreteness, turns out to be attractive and worthwhile \cite{KumarWalia:2022ddq,Ashtekar:2002sn,Boehmer:2007ket,Peltola:2008pa,Ashtekar:2005qt}.  Most of the regular black holes closely connected to a potential theory of quantum gravity are not derived from but are \emph{inspired} by quantum gravity. They make up applicable phenomenological models but make their physical justification less straightforward \cite{Ashtekar:2004eh}.

Peltola and Kunstatter \cite{Peltola:2009jm}, following earlier works \cite{Ashtekar:2005qt,Modesto:2005zm,Modesto:2008im,Modesto:2006mx,Boehmer:2007ket,Boehmer:2008fz,Campiglia:2007pb,Gambini:2008dy}, used the effective field theory technique to the symmetry reduced black hole Hamiltonian reported in terms of the two configuration space variables: the areal radius $r$ and a function $\rho$ obtained from the conformal mode of the metric. The effective field equations that arise are readily resolvable and guide to the following black hole exterior metric
 \cite{Peltola:2008pa,Peltola:2009jm}
\begin{eqnarray}\label{metric}
	ds^{2}&=&-\left(\sqrt{1-\frac{l^2}{y^2}}-\frac{2M}{y}\right)dt^{2}+\frac{\left(1-\frac{l^2}{y^2}\right)^{-1}}{\left(\sqrt{1-\frac{l^2}{y^2}}-\frac{2M}{y}\right)}dy^{2} \nonumber\\
	&+& y^2 (d\theta^{2}+\sin^{2}\theta d\phi^{2}),
\end{eqnarray}
Thereby the dynamical field equations in the partially polymerized theory admit the above static and spherically symmetric black holes  \cite{Peltola:2008pa,Peltola:2009jm} 
with $t$ being the Schwarzschild time, $r$ being the areal radius and $l$ being the polymer length scale.
 The solution is asymptotically flat at $y\to \infty$.  One of the most striking features of this four-dimensional quantum-corrected black hole metric (\ref{metric}) is that it has a single horizon at $y\equiv y_+=\sqrt{4M^2+l^2}$. 
 \begin{figure*} 
	\begin{centering}
		\begin{tabular}{p{9.5cm} p{9.5cm}}
		    \includegraphics[scale=0.7]{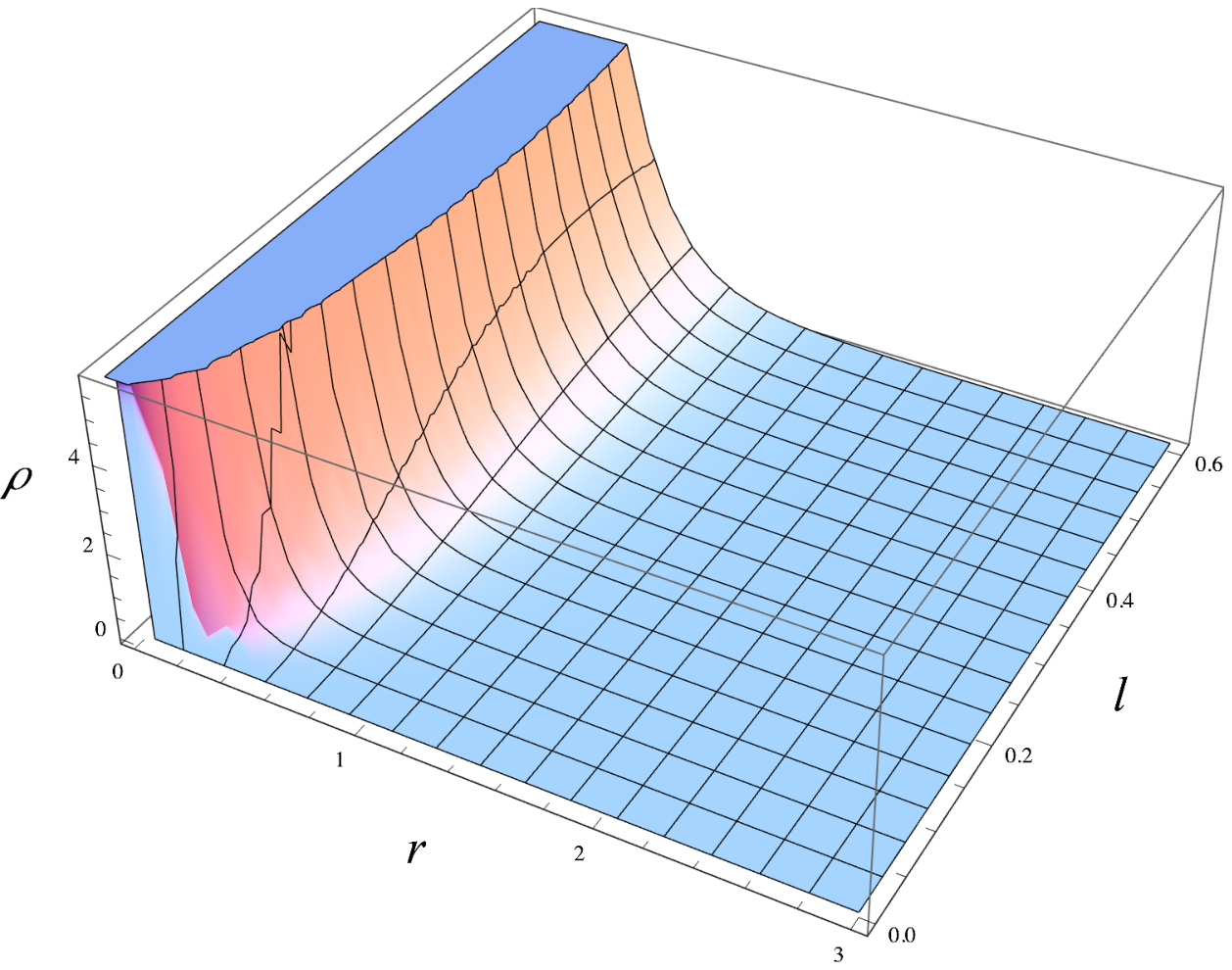}&
		    \includegraphics[scale=0.7]{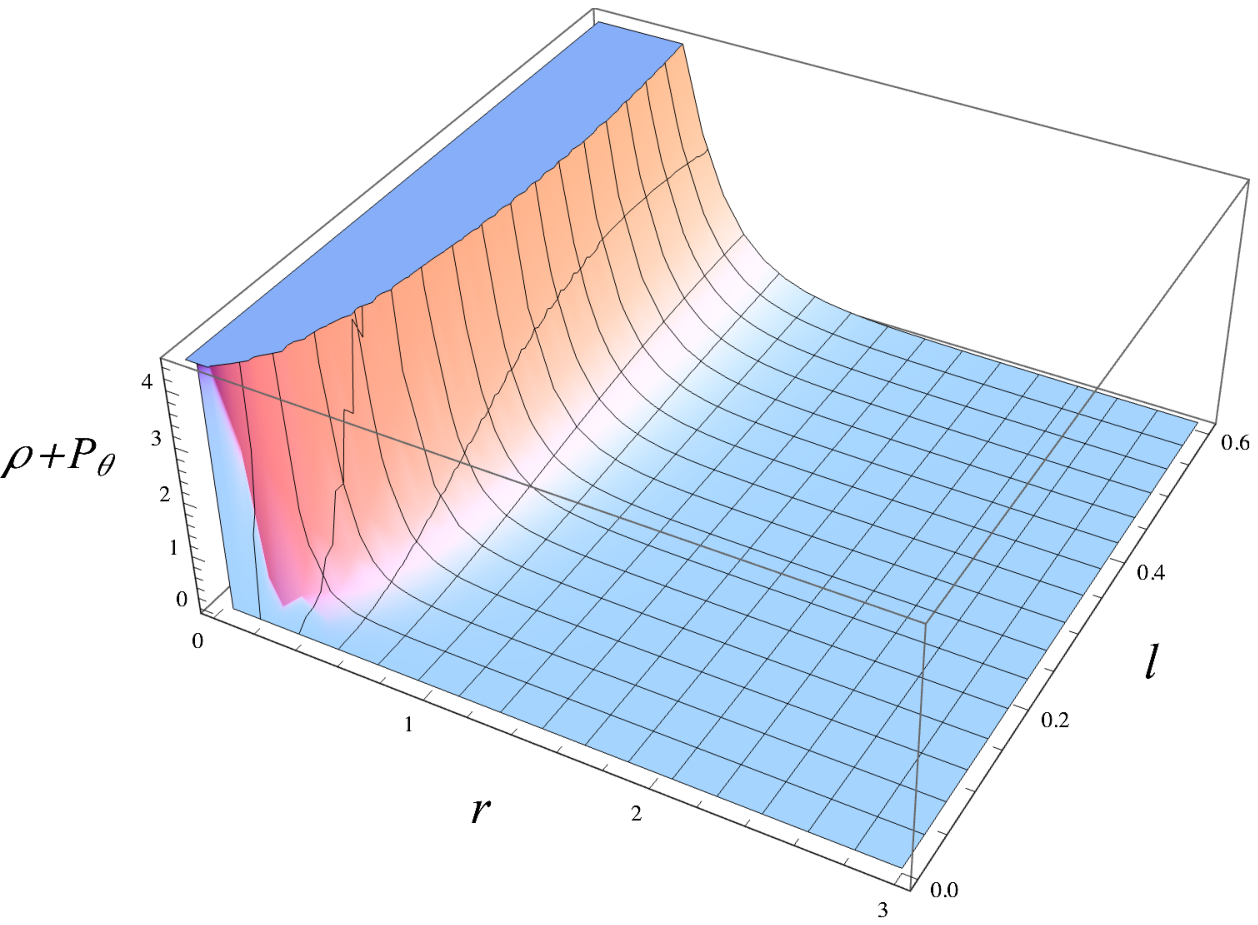}
			\end{tabular}
	\end{centering}
	\caption{Behaviour of $\rho$,  $\rho+P_{\theta}$ for different values of the parameter $l$. }\label{energycond}
\end{figure*}
Thus we have several effective polymerized spherically symmetric black hole solutions.  However, the spherical black hole models can hardly be tested by astrophysical observations, as the black hole spin, i.e., rotating black holes, typically found in nature,  plays a fundamental role in any astrophysical process. Further, without rotating black hole models, LQG substantially hinders the progress of testing LQG theory from observations. The Kerr metric \cite{Kerr:1963ud} is   the most important black hole solutions of general relativity that can arise from gravitational collapses of massive star. 
Also, it turns out that when applied to other models in LQG, the revised Newman-Janis works quite well in generating rotating metrics starting with their non-rotating  seed metrics \cite{Brahma:2020eos,Liu:2020ola, Chen:2022nix,Ghosh:2014pba,Ghosh:2015ovj,Kumar:2020hgm,Kumar:2020owy,Kumar:2021cyl,Islam:2021dyk}.  This encouraged us to pursue  rotating or axisymmetric generalization of the spherical metric (\ref{metric}) or finding a Kerr-like metric, namely, a LQG motivated rotating black hole (LMRBH) metric suitable for test with astrophysical observations. This is achieved by by starting with a spherical 
LQG black hole  (\ref{metric}) as a seed metric, we construct a rotating spacetime using the revised Newman-Janis
algorithm. We discuss the various black hole properties including the horizon structure, and  construct the Penrose diagrams and the embedding diagrams .

The organization of this paper is as tails. We begin in Sec.~\ref{sec2} with the building of the rotating counterpart of the spherical seed metric (\ref{metric2}), namely, the LMRBH metric. Before this, we also discuss generic features of the  LQG polymarized single horizon spherical black hole in the same section. In Sec.~\ref{sec3}, we discuss the horizon structure and phase space of the LMRBH spacetimes and we present its peculiar properties, e.g, the LMRBH spacetimes, depending on the values of parameters $a$ and $l$, can admit up to three horizons. Other is the possibility that the black hole spin parameter $a$ may surpass the black hole mass parameter $M$. An extreme black hole in LMRBH will have $a>M$; we provide numerical illustrations in Sec.~\ref{sec3}. In Sec.~\ref{sec4}, we investigate the the global structure of the LMRBH via Penrose diagrams of LMRBH and the extended spacetime. We also  discuss here the embedding diagrams of the equatorial plane of LMRBH spacetime for the selected regions of the parameter space. Finally, we outline our main results in  Sec.~\ref{sec5}.

\section{LQG motivated Polymerized Rotating Black Holes}\label{sec2} We can do away with the coordinate singularity at $y=l$ for the metric (\ref{metric}), by using the transformation $y=\sqrt{r^2+l^2}$; then the metric takes the form
\begin{eqnarray}\label{metric2}
ds^{2}&=& -\left(\frac{r-2M}{\sqrt{r^2+l^2}}\right)dt^{2}+\frac{1}{\left(\frac{r-2M}{\sqrt{r^2+l^2}}\right)}dr^{2}\nonumber\\
&+&(r^2+l^2)(d\theta^{2}+\sin^{2}\theta d\phi^{2}).
\end{eqnarray}
Here, the radial coordinate $r$ assumes the full range $0\leq r\leq \infty$. This metric irrespective of the value of $l$ always represents a black hole and has a single horizon, namely event horizon located at fixed radial coordinate, $r=2M$. 
\begin{figure*}
\begin{center}
		    \includegraphics[scale=1.0]{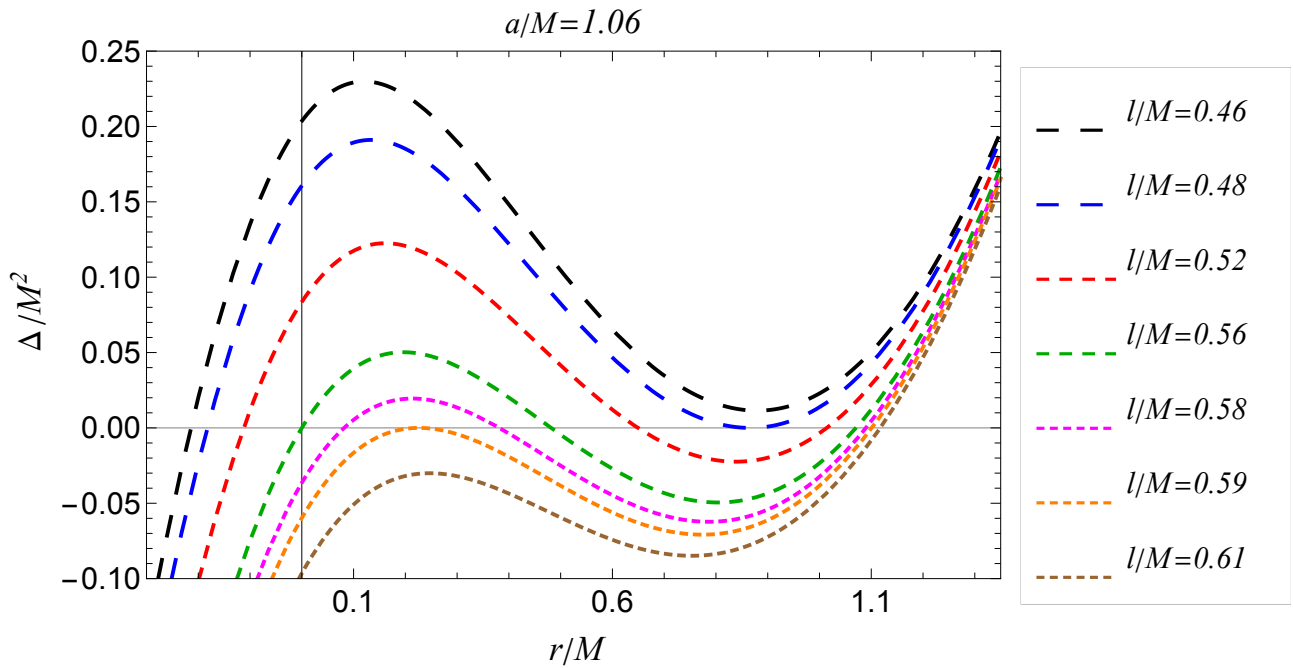}
\end{center}
	\caption{Plot showing behaviour of horizons ($\Delta (r)$ vs $r$) for fixed $a/M$ and varying $l/M$; the points where $\Delta (r)=0$ are the locations of horizons.}\label{horizon1}
\end{figure*}
It is necessary to investigate the behaviour of scalar invariants in order to assess the regularity of solution (\ref{metric2}). For this metric,  the invariants Ricci scalar $R = R_{ab} R^{ab}$, ($R_{ab}$ Ricci tensor) and Kretschmann scalar $K = R_{abcd} R^{abcd}$ ($R_{abcd}$ Reimann tensor), respectively, assume the form
\begin{eqnarray}
R &=& \frac{1}{\left(l^2+r^2\right)^{5/2}}\left[l^2 (6 M-5 r)+2 \left(l^2+r^2\right)^{3/2}-2 r^3\right]\\
K &=& \frac{1}{\left(l^2+r^2\right)^5}[ 12 r^4 \left(l^2+4 M^2\right)+r^2 \left(41 l^4-48 l^2 M^2\right) \nonumber\\ &&
+36 l^4 M^2+56 l^2 M r^3-60 l^4 M r+4 l^6-16 M r^5+8 r^6 \nonumber\\ &&-8 r^2 \left(l^2+r^2\right)^{3/2} (r-2 M)]
\end{eqnarray}
Hence, these invariants are everywhere finite, and vanish rapidly at far distances from the black hole. As a result, the polymerized black hole metric describes a globally regular spacetime.
\subsection{Energy conditions}
From the Einstein field equation, we have $G^{\mu}_{\nu}=T^{\mu}_{\nu}$. Here, $T^{\mu}_{\nu}$ is the energy momentum tensor associated with matter, which reads
\begin{equation}
    T^{\mu}_{\nu}=\mbox{diag}(-\rho, P_r, P_{\theta}, P_{\phi}).
\end{equation}

We use Hawking-Ellis' approach to assess the  status of the various energy conditions \cite{Hawking:1973uf,Ghosh:2008zza,Kothawala:2004fy}. The weak energy condition (WEC) demands $T_{\mu\nu}\xi^{\mu}\xi^{\nu}\geq 0$ for any timelike vector
$\xi^{\mu}$. The WEC requires
\begin{equation}\label{1a}
\rho\geq 0;  ~~ ~\rho + P_i \geq 0,~~~ (i=r,\; \theta,\; \phi)
\end{equation}
which guarantees that the energy density as measured by any local observer is non-negative \cite{Hawking:1973uf}. The strong energy condition (SEC), in addition to Eq. (\ref{1a}) requires
\begin{equation}\label{sec}
\rho + P_r + 2P_{\theta} \geq 0.
\end{equation}
In our case, i.e. for the metric (\ref{metric2}), we have
\begin{eqnarray}
\rho &=& \frac{4 l^2 M+\left(l^2+r^2\right)^{3/2}-3 l^2 r-r^3}{\left(l^2+r^2\right)^{5/2}}\label{rho},\\
\rho+P_{r} &=& -\frac{2 l^2 (r-2 M)}{\left(l^2+r^2\right)^{5/2}}\label{1aa},\\
\rho+P_{\theta} &=&  \rho+P_{\phi} = \frac{l^2 \left(2 \sqrt{l^2+r^2}+6 M-5 r\right)}{2 \left(l^2+r^2\right)^{5/2}}\nonumber \\ && + \frac{2 r^2 \left(\sqrt{l^2+r^2}-r\right)}{2 \left(l^2+r^2\right)^{5/2}} \label{2aa},\\
\rho+P_{r}+2~P_{\theta} &=& -\frac{l^2 (r-2 M)}{\left(l^2+r^2\right)^{5/2}}.
\end{eqnarray}
First we note that $\rho+P_r > 0$ for $r<2M$. The behaviour of $\rho$ and $\rho+P_{\theta}$ is depicted in Fig. \ref{energycond} which are positive for most of the values of $r$. Thus, the WEC is mostly satisfied when $r<2M$. The violation can be very small, depending on the value of $l$, as seen from Fig. \ref{energycond}. Similarly, $\rho+P_{r}+2P_{\theta}>0$ for $r<2M$ and SEC is obeyed for most values of $r<2M$, and violation, as in the case of WEC, is minimal.

The solution (\ref{metric2}) will be the spherical \emph{seed} metric to generate the rotating LMRBH metric by a non-complexification technique \cite{Azreg-Ainou:2014pra,Azreg-Ainou:2014aqa}.
\begin{figure*} 
\begin{center}
\includegraphics[scale=1.0]{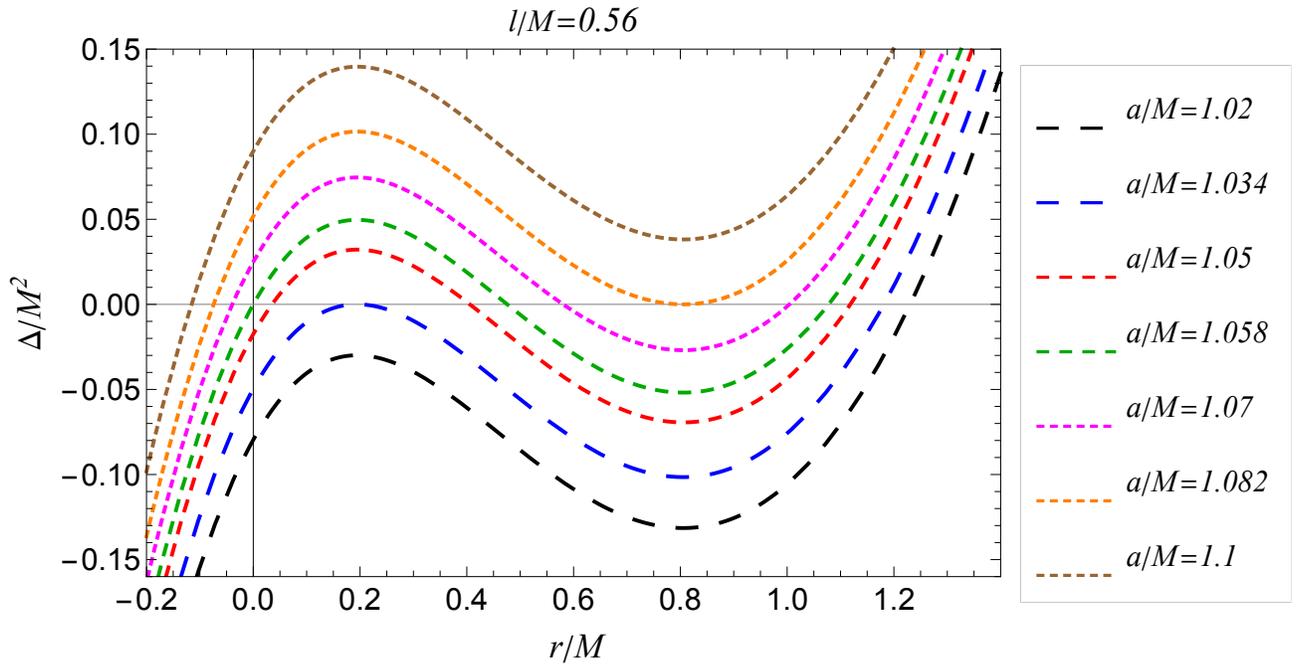}
\end{center}
	\caption{Plot showing behaviour of horizons ($\Delta (r)$ vs $r$) for fixed $l/M$ and varying $a/M$; the points where $\Delta (r)=0$ are the locations of horizons.}\label{horizon2}
\end{figure*}
\subsection{Rotating metric}
The Newman$-$Janis algorithm (NJA) has been commonly used to generate rotating black hole solutions from their non-rotating or spherical counterparts \cite{Newman:1965tw}. While this algorithm was developed within general relativity, it has been more recently applied to non-rotating solutions in modified gravity theories \cite{Johannsen:2011dh,Jusufi:2019caq,Ghosh:2014hea,Moffat:2014aja}. Exercising the Newman$-$Janis algorithm to an arbitrary non-general relativity spherically symmetric solution can introduce pathologies in the resulting axially symmetric metric \cite{Hansen:2013owa}. Hence, we derive the rotating LMRBH black hole metric, using the modified NJA developed by Azreg-A\"inou's - a non-complexification procedure\cite{Azreg-Ainou:2014pra,Azreg-Ainou:2014aqa}. The researchers have successfully applied it for generating imperfect fluid rotating solutions in the Boyer$-$Lindquist coordinates from spherically symmetric static black holes \cite{Kumar:2022fqo,Afrin:2021wlj,Ghosh:2022jfi, Walia:2021emv,Kumar:2020owy}, including some other rotating LQG metrics. 
We derived the LQG-motivated rotating black hole (LMRBH) metric, starting from the spherical  metric (\ref{metric2}), via the modified Newman\(-\)Janis algorithm \cite{Azreg-Ainou:2014pra,Azreg-Ainou:2014aqa}, which in Boyer\(-\)Lindiquist coordinates reads
\begin{eqnarray}\label{metric3}
ds^2 &=& -\left[1-\frac{2M(r)\sqrt{r^2+l^2}}{\rho^2}\right] dt^2+ \frac{\rho^2}{\Delta} dr^2 +\rho^2 d\theta^2  \nonumber\\ && - 4aM(r) \sqrt{r^2+l^2} \sin^2\theta dtd\phi+ \frac{\mathcal{A}\sin^2\theta~}{\rho^2} d\phi^2
\end{eqnarray}
where
\begin{eqnarray}
M(r) &=& M-\frac{r-\sqrt{r^2+l^2}}{2} ~~~~~
\rho^2 = r^2+l^2 +a^2\cos^2\theta,\nonumber \\
\Delta &=& r^2+l^2 +a^2 -2 M(r) \sqrt{r^2+l^2},\nonumber \\
\mathcal{A} &=& (r^2+l^2 +a^2)^2-a^2 \Delta \sin^2\theta.
\end{eqnarray}

\begin{figure*}
\begin{center}
	\includegraphics[scale=1.0]{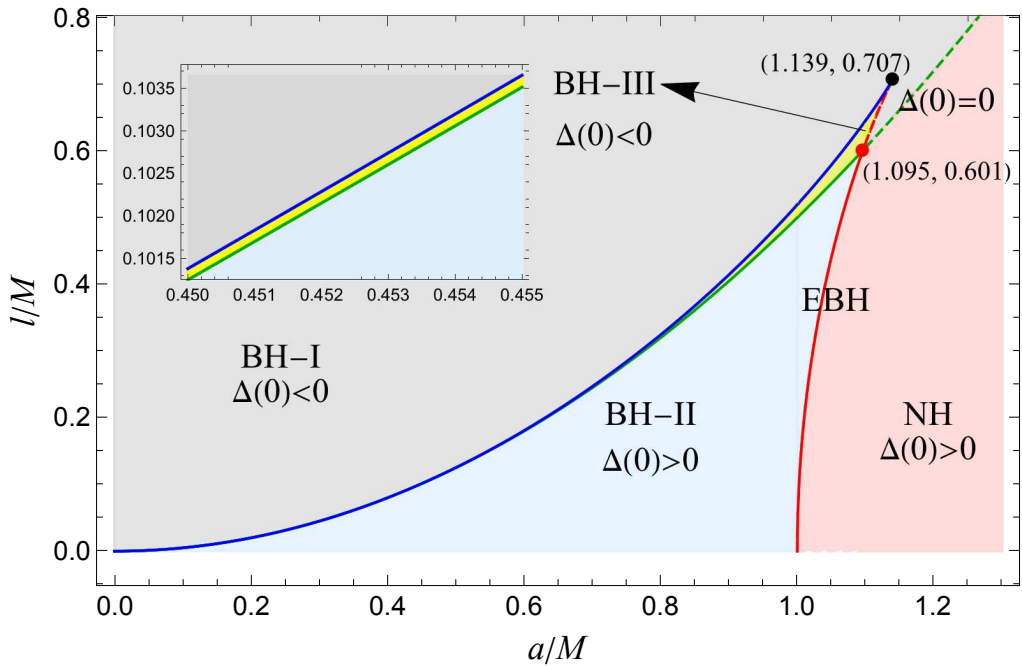}
	\caption{Parameter space ($M,a,l$) for LMRBH metric (\ref{metric3}). We spell acronyms and details in the text. The red transition line (EBH) corresponds to the extremal black holes with degenerate horizons, where the outer two horizons merge. The blue transition line also corresponds to the black hole with degenerate horizons, but unlike the red line, the inner two horizons merge. For the  green line, $\Delta(r)=0$ admits two positive and $r=0$ root while for dashed green line, we have $r=0$ as the only root.}\label{parameter}
\end{center}	
\end{figure*}

The LMRBH metric (\ref{metric3}) reverts to the standard Kerr black hole metric in the limit $l \to 0$, and to the spherical metric (\ref{metric2}) when $a=0$. The LMRBH spacetime, like Kerr spacetime, admits two killing vectors $\eta^{\mu}_{(t)}$ and $\eta^{\mu}_{(\phi)}$ being invariant under $t$ and $\phi$ transformations.

The horizons are surface defined by $g^{rr}=0=\Delta$ and the radii of horizons are zeroes of 
\begin{equation}\label{delta}
(r-2M)\sqrt{r^2+l^2}+a^2=0.   
\end{equation}
Eq. (\ref{delta}), for special case $l=0$, reduces to $r^2-2Mr+a^2=0$ which admits solutions $r_\pm=M\pm \sqrt{M^2-a^2}$, which are radii of Kerr black hole.
Analysis of $\Delta(r)$ reveals that $\Delta(r)$ has a local maxima at $r=\left(M-\sqrt{M^2-2l^2}\right)/2$ and a local minima at $r=(M+\sqrt{M^2-2l^2})/2$ when $l<M/\sqrt{2}$ and is strictly increasing for $l\geq M/\sqrt{2}$. Also $\Delta(r) \to \pm\infty$ as $r \to \pm\infty$. Therefore, $\Delta(r)=0$ will necessarily have at least one root and can have maximum of three roots depending on the values of $a$ and $l$, with possibility of one of the roots being negative (cf. Figures \ref{horizon1} and \ref{horizon2}). We name the roots $r_1$, $r_2$, $r_3$ with $r_3 \le r_2 \le r_1$. The largest root, $r_1$ is always an event horizon. $r_2 \le r_1$, if it exists, is always a Cauchy horizon. While $r_3$ is an additional horizon inside the Cauchy horizon and when $r_3<0$, it corresponds to a horizon inside $r=0$ surface. The parameter space ($M,a,l$) for LMRBH is depicted in  Fig.~\ref{parameter}. Note that  positive roots of $\Delta(r)=0$ correspond to the horizons of LMRBH with the larger/smaller root corresponding to the event /Cauchy horizon.  
\begin{table*}
\begin{tabular}{|l| l| l | l |l | l|}  
\hline
\multicolumn{1}{|c|}{Lines and Dots in Fig.~\ref{parameter}}&
\multicolumn{1}{|c|}{Transition}& 
\multicolumn{2}{|c|}{Critical value of $l$ for given $a$} & 
\multicolumn{2}{|c|}{Critical value of $a$ for given $l$}\\
 &  &  ~~~$a/M$~~~ & ~~~$l_c/M$~~~ &~~~ $l/M$~~~ &~~~ $a_c/M$ \\
\hline\hline
\multirow{2}{*}{Red line} & BH-II to NH & 
 1.04 &  0.566629 &  0.2 &  1.01005 \\
&  &  1.07 &  0.60462 &  0.4 &  1.04091 \\
\hline

\multirow{2}{*}{Dashed Red  line} & BH-III to BH-I & 
  1.1 &   0.645459 &   0.64 &   1.11044 \\
&  &   1.12 &  0.674884 &  0.68 &  1.12694 \\
\hline
\multirow{2}{*}{Green line} & BH-III to BH-II & 
 0.5 & 0.125 & 0.3 & 0.774597 \\
&  &  0.7 &  0.245 &  0.55 & 1.04880 \\
\hline
\multirow{2}{*}{Dashed Green  line} & BH-I to NH & 
 1.12 & 0.6272 & 0.65 & 1.14017\\
&  &  1.2 &  0.72 &  0.7 & 1.18321 \\
\hline
\multirow{2}{*}{Blue line} & BH-III to BH-I & 
 0.6 &  0.180746 &  0.4 &  0.884923 \\
 &  &  0.8 & 0.32442 & 0.6 & 1.06646 \\
\hline
\multirow{1}{*}{Red Dot} & Surrounded by BH-III, 
BH-I and NH  & 
 1.095 & 0.601 & 0.601 & 1.095 \\
\hline
\multirow{1}{*}{Black Dot} & Surrounded by BH-III, BH-I & 
 1.139 & 0.7076 & 0.7076 & 1.139 \\
\hline
\end{tabular}
\caption{ Table summarizing the  critical values of parameters ($a_c/M,l_c/M$) for different lines and dots in parameter space in Fig.~\ref{parameter}. }\label{table1}
\end{table*}

It turns out that, for a given $a\;(l)$, there is a critical value of $l (a)$, $l_c (a_c)$ (on the red line), such that $\Delta(r)=0$ has a double root which corresponds to an extremal LMRBH with degenerate horizons. Further, for $a<a_c (l>l_c)$, $\Delta(r)=0$  admit two simple positive zeros corresponding LMRBH with Cauchy and event horizons (BH-II), while for $a>a_c (l<l_c)$, $\Delta(r)=0$  admit not positive zeros and correspond to no-horizon (NH) spacetimes.  Similarly, for given  $a\;(l)$, we can find critical value of $l (a)$, $l_c (a_c)$ (see Table~ \ref{table1}) on other transition lines and dots (Fig.~\ref{parameter}). 

\section{Phase Diagram}\label{sec3}
The range of possibilities with each spacetime structure corresponding to a particular region of the parameter space are defined as follows. 
We summarise the range of cases with a ``phase diagram'' depicted in Fig.~\ref{parameter}. We use the following terminology for the parameter space under consideration for different spacetime structures (cf. Fig.~\ref{parameter}). \\

\begin{figure*}
	\begin{centering}
		\begin{tabular}{p{9.5cm} p{9.5cm}}
		    \includegraphics[scale=0.75]{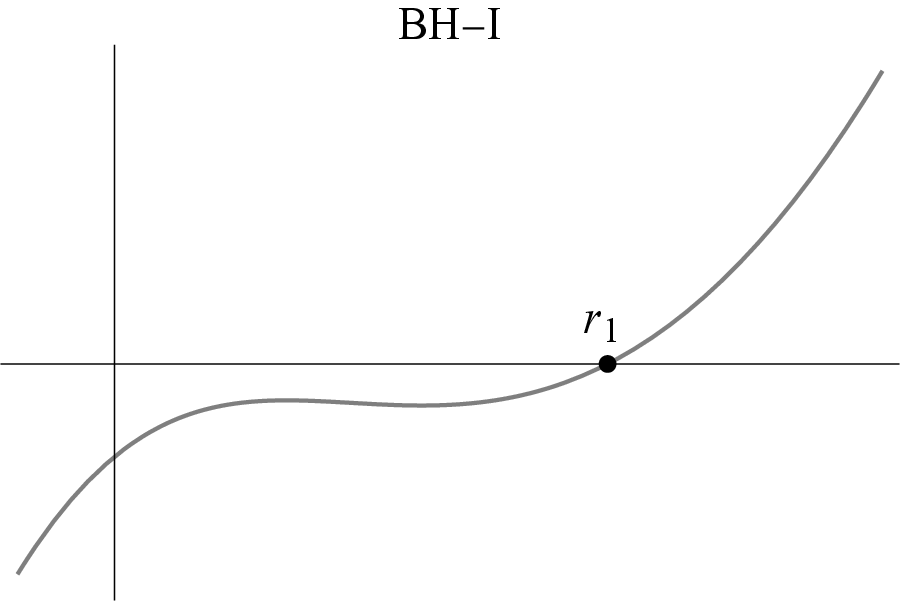}&
		    \includegraphics[scale=0.75]{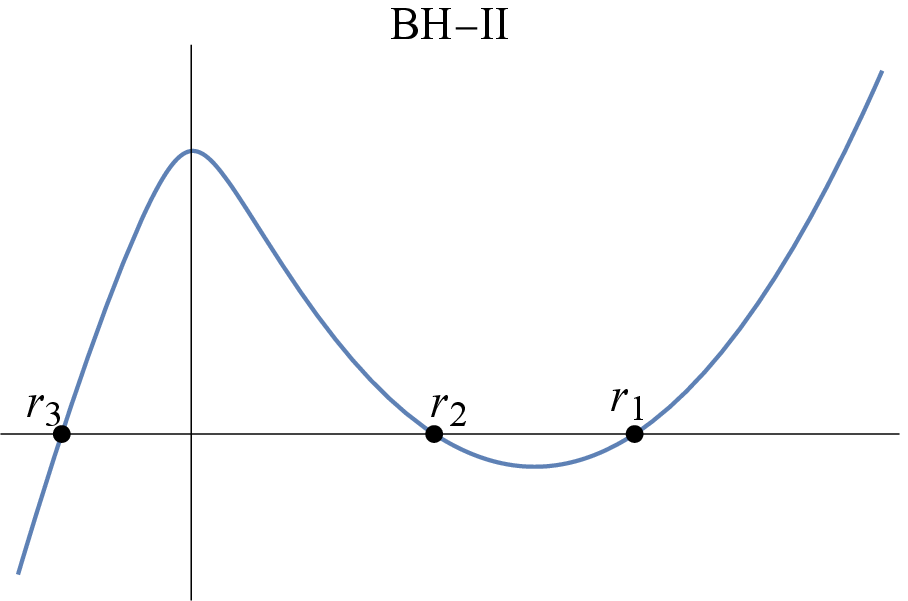}\\
		    \includegraphics[scale=0.4]{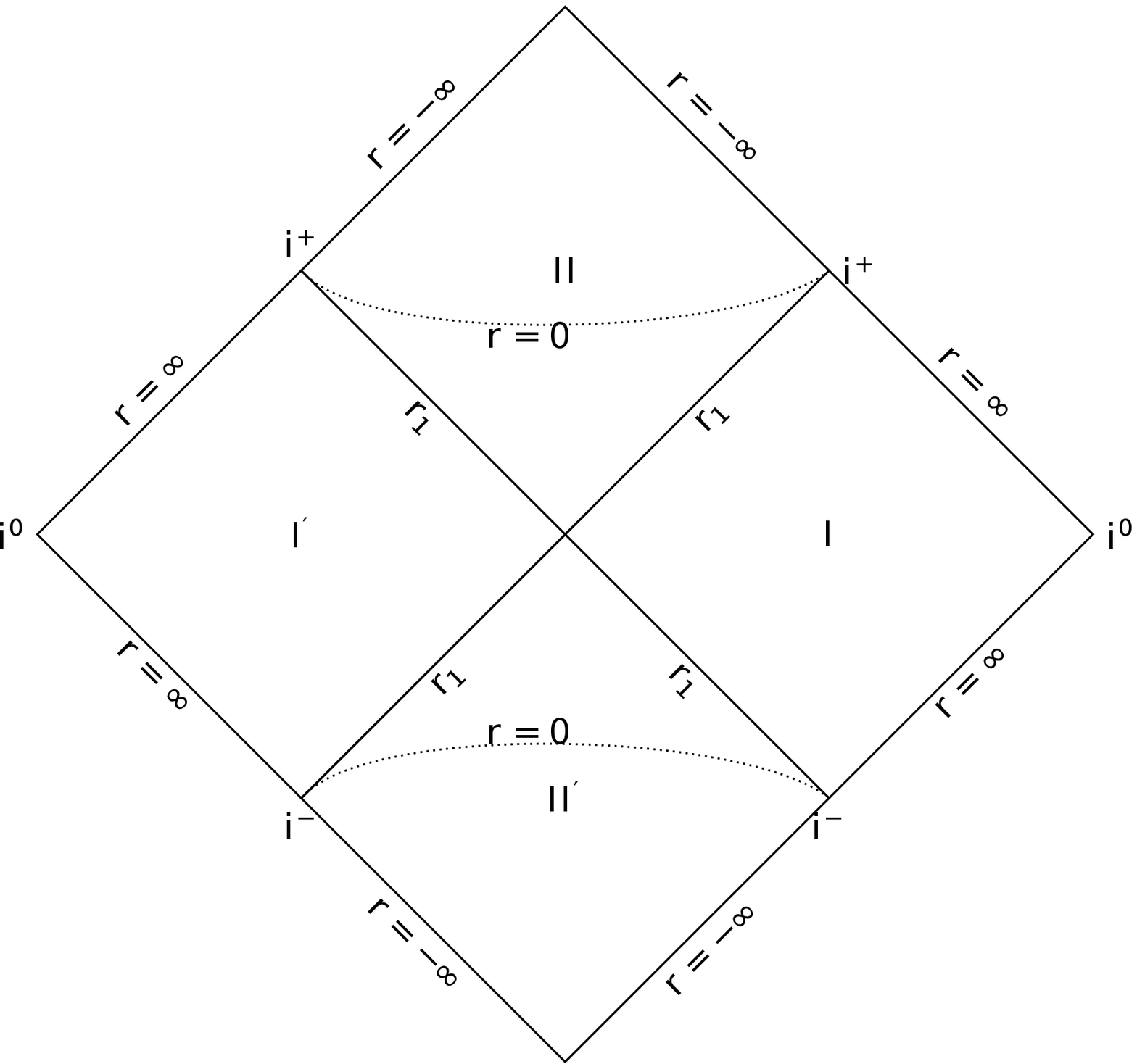}&
		     \includegraphics[scale=0.45]{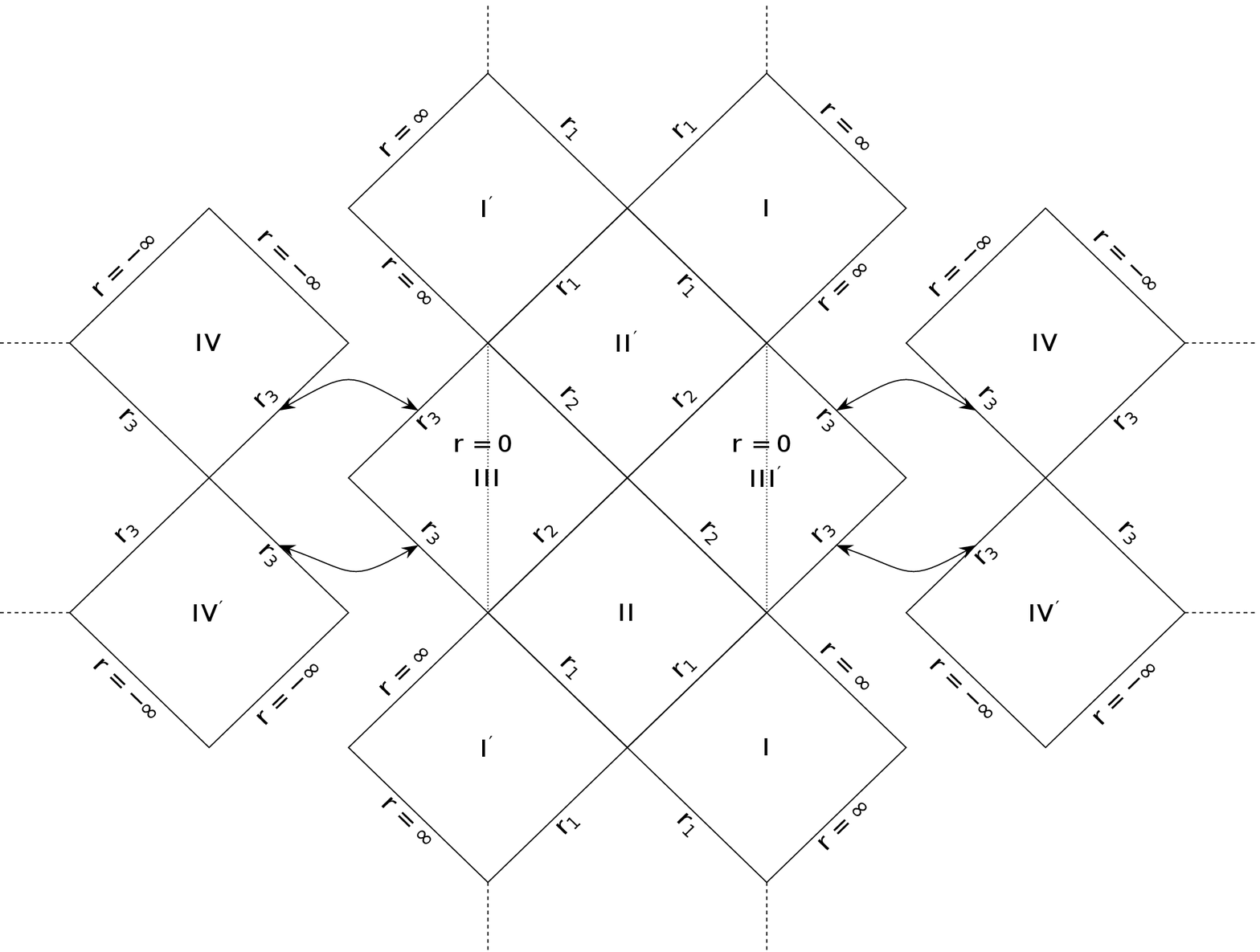}
			\end{tabular}
	\end{centering}
	\caption{Plot showing $\Delta(r)$  in the parameter space (Fig. \ref{parameter}) (a) BH-I, where $\Delta(r)=0$ admits only one positive root $r_1=$ corresponding to event horizon (Top left), (b) BH-II where $\Delta(r)=0$ admits two distinct positive roots $r_1$ (event horizon) and $r_2$ (Cauchy horizon), and one negative root $r_3$ corresponding to additional third horizon (Top right).
	Penrose diagrams of LMRBH in the parameter space $(M,a,l)$: (a) BH-I  (Bottom left) and (b) BH-II (Bottom right). They will identify the surfaces connected by the arrows as the same surfaces.}\label{pen4}
\end{figure*}

\begin{figure*} 
	\begin{centering}
		\begin{tabular}{p{9.5cm} p{9.5cm}}
		    \includegraphics[scale=0.75]{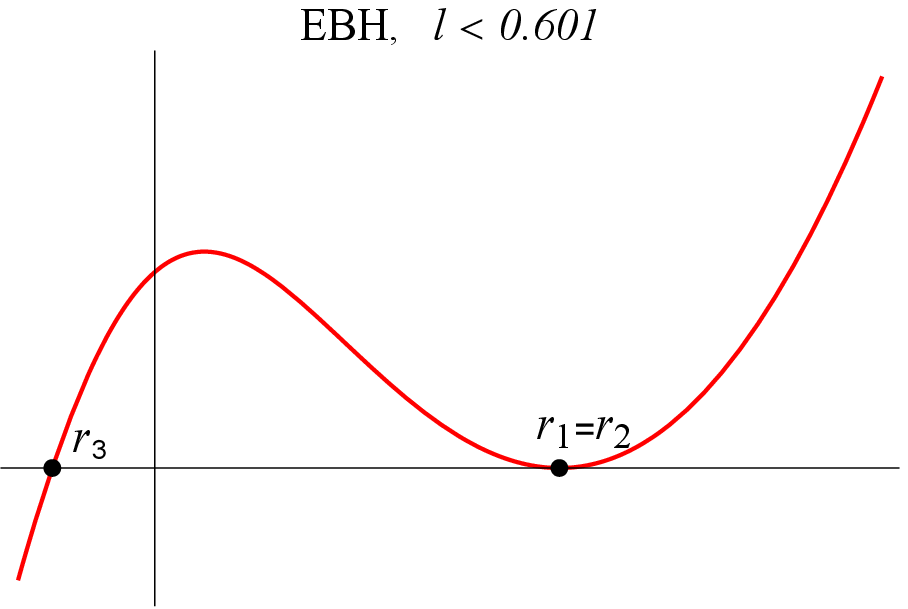}&
		    \includegraphics[scale=0.75]{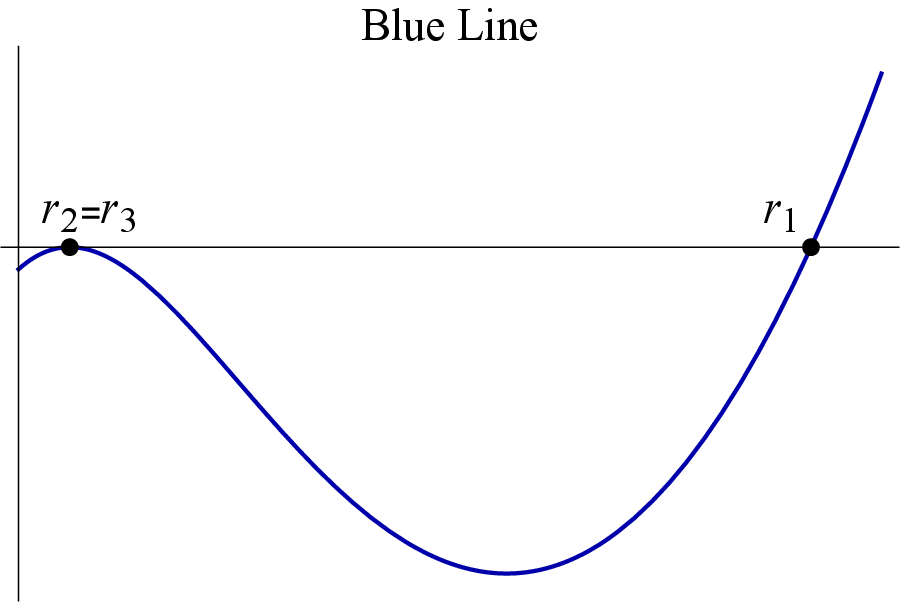}\\
		    \includegraphics[scale=0.52]{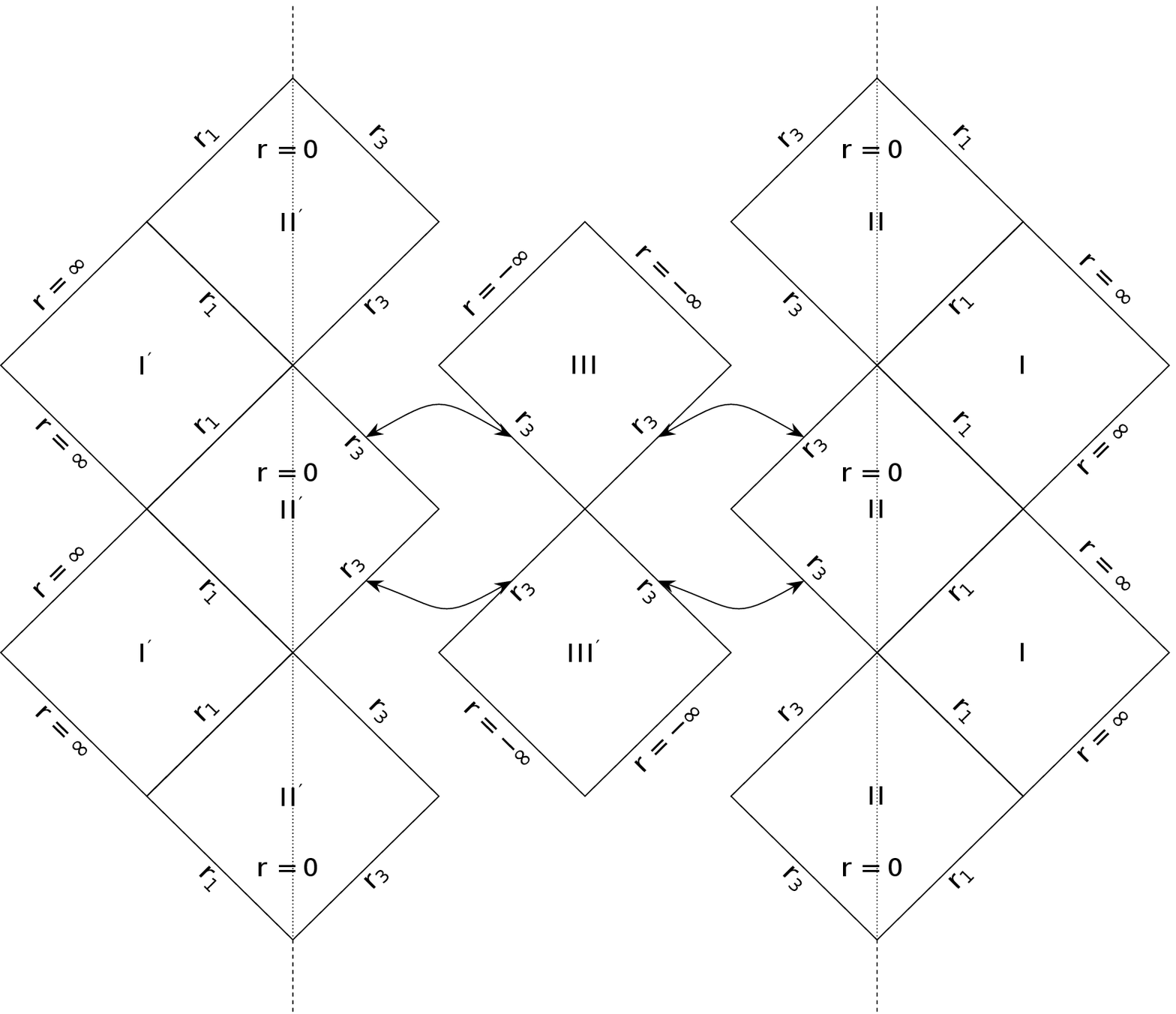}&
		     \includegraphics[scale=0.5]{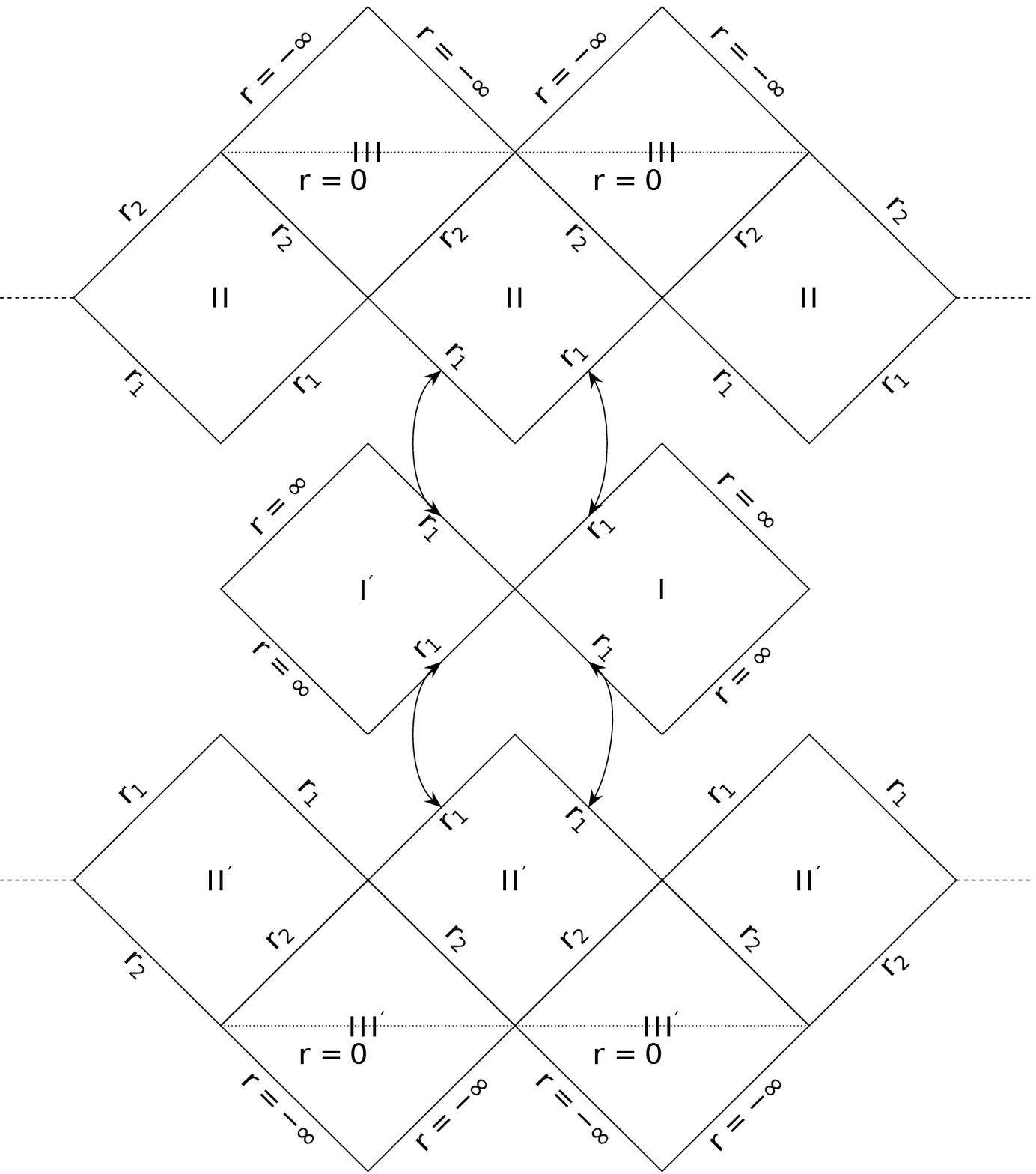}
			\end{tabular}
	\end{centering}
	\caption{Plot showing $\Delta(r)$ in the parameter space (Fig. \ref{parameter}) (a) EBH $l<0.601M$, where $\Delta(r)=0$ admits two positive equal roots $r_1=r_2$, corresponding to degenerate horizons and one negative root $r_3$, corresponding to the additional third horizon (Top left), (b) Blue line, where $\Delta(r)=0$ admits two positive equal roots $r_2=r_3$ corresponding to degenerate horizons and one larger root $r_1$ corresponding to event horizon(Top right).
	Penrose-Carter diagrams of the parameter space $(M,a,l)$ showing the region (a)  EBH $l<0.601M$ (Bottom left) and (b) Blue line (Bottom right). The surfaces connected by the arrows are to be identified as the same
 surfaces.}\label{pen5}
\end{figure*}

\begin{figure*} 
	\begin{centering}
		\begin{tabular}{p{6cm} p{6cm}  p{6cm}}
		    \includegraphics[scale=0.55]{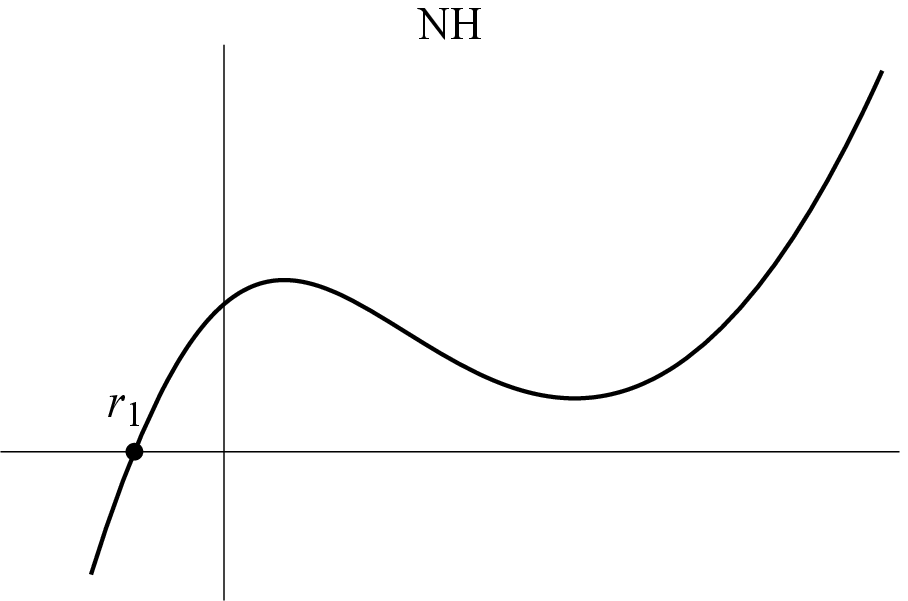}&
		    \includegraphics[scale=0.55]{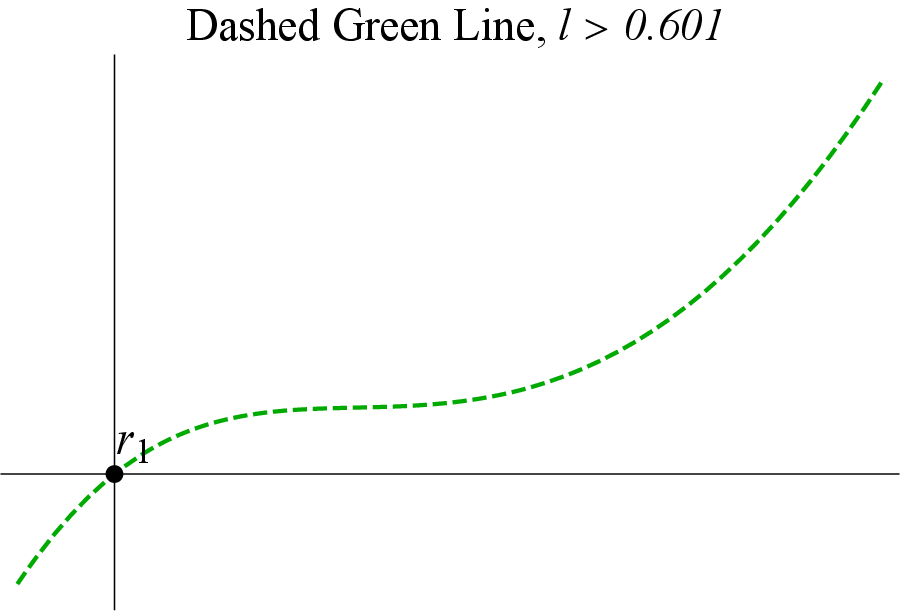}&
		    \includegraphics[scale=0.55]{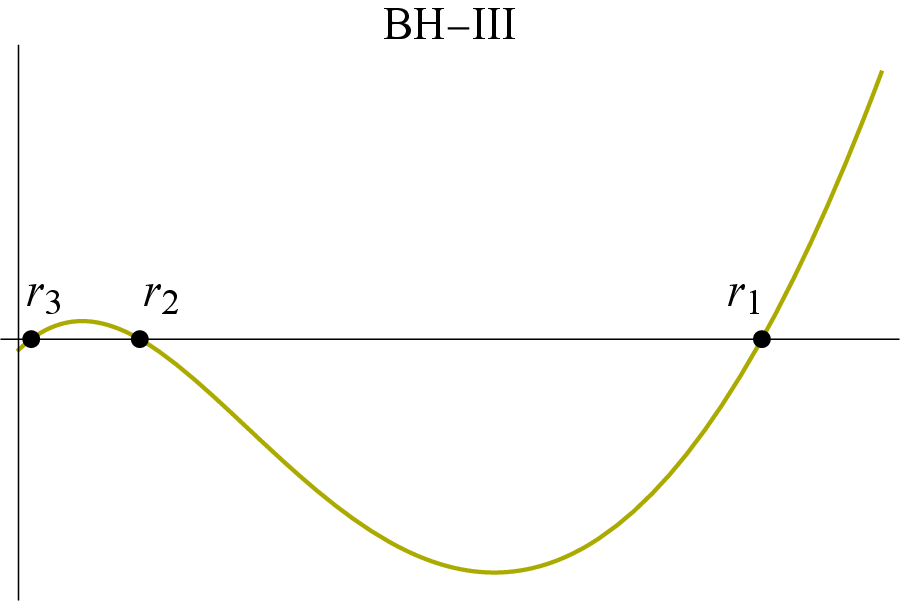}
			\end{tabular}
	\end{centering}
	\caption{Plot showing $\Delta(r)$ in the region of parameter space (Fig. \ref{parameter}) (a) NH, where $\Delta(r)=0$ admits only one negative root $r_1$, corresponding to horizon inside $r=0$ surface (Left), (b) Dashed green line, $l>0.601M$ where $\Delta(r)=0$ admits only one root $r_1=0$, showing that transition surface $r=0$ is null (Middle), (c) BH-III, where $\Delta(r)=0$ admits three distinct positive roots $r_1,r_2,r_3$ with $r_1$ and $r_2$, respectively, corresponding to event and Cauchy horizon while $r_3$ is the additional third horizon (Right). The Penrose diagrams for NH and dashed green line, $l>0.601M$ are same as in Fig. \ref{pen4} (left) except that the surface $r=0$ lies in the regions I and I$'$ and is timelike for NH; while for the case of dashed green line, $l>0.601M$, $r=r_1$ is replaced by the null surface $r=0$. The Penrose diagram for BH-III is the same as in Fig. \ref{pen4} (right), except that the surface $r=0$ surface lies in the regions IV and IV$'$ and is spacelike.}\label{hor1}
\end{figure*}

\begin{figure*} 
	\begin{centering}
		\begin{tabular}{p{6cm} p{6cm}  p{6cm}}
		    \includegraphics[scale=0.55]{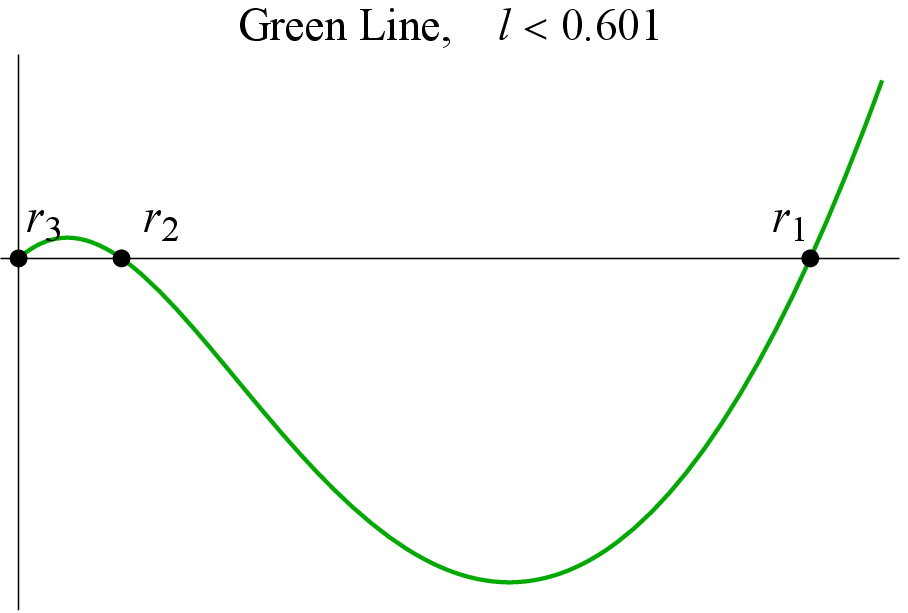}&
		    \includegraphics[scale=0.55]{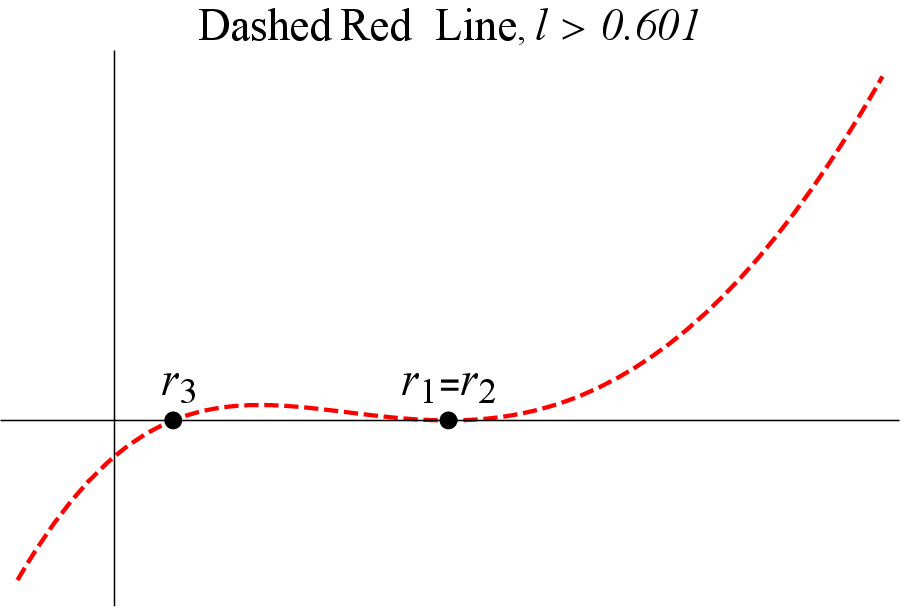}&
		    \includegraphics[scale=0.55]{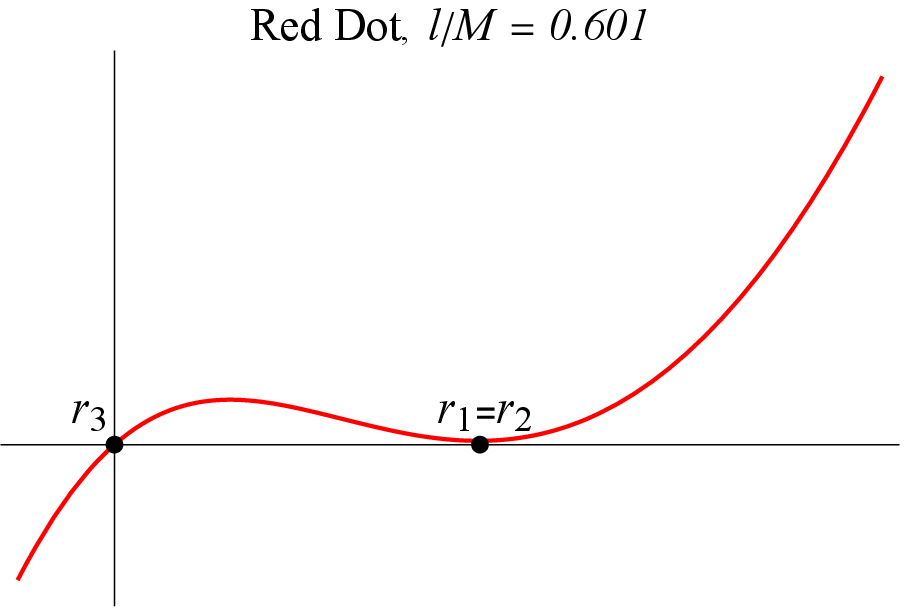}
			\end{tabular}
	\end{centering}
\caption{Plot showing $\Delta(r)$ in the region of parameter space (Fig. \ref{parameter}) (a) Green line, $l<0.601M$ where $\Delta(r)=0$ admits two distinct positive roots $r_1$ (event horizon) and $r_2$ (Cauchy horizon) and one additional root at $r_3=0$ (null transition surface) (Left), (b) Dashed red line $l>0.601M$, where $\Delta(r)=0$ admits two positive equal roots $r_1=r_2$, corresponding to degenerate horizons and one smaller positive root $r_3$ corresponding to the additional third horizon (Middle), (c) Red dot $a=1.095M, l=0.601M$, where $\Delta(r)=0$ admits two positive equal roots $r_1=r_2$, corresponding to degenerate horizons and one root at $r_3=0$ (null transition surface) (Right). The Penrose diagram for green line, $l<0.601M$ is same as in Fig. \ref{pen4} (right) except that $r=r_3$ is replaced by the null surface $r=0$. The Penrose diagrams for dashed red line and red dot are the same as in Fig. \ref{pen5} (left) except that the surface $r=0$ lies in the regions III and III$'$ and is spacelike for dashed red line; while for the case of red dot, $r=r_3$ is replaced by the null surface $r=0$.  }\label{hor2}
\end{figure*}

\begin{itemize}
\item \textbf{BH-I}:
In this region of parameter space (The grey region in Fig.~\ref{parameter}) $\Delta(r)=0$ has only one positive root (cf. Fig.~\ref{pen4}) corresponding to the single event horizon similar to the spherical black hole (\ref{metric}), and we refer to this as single horizon LMRBH .
\item \textbf{BH-II}:
 $\Delta(r)=0$ has two positive roots $(r_2 < r_1)$ in this region  and one negative root ($r_3$) (light blue region in Fig.~\ref{parameter}). The former case corresponds to the Cauchy ($r_2$) and event horizon ($r_1$) (cf. Fig.~\ref{pen4}).  This region is the most physically relevant one for considering rotating black holes and we refer to this as \textit{generic} LMRBH. 
\item \textbf{BH-III}:
In the yellow region in Fig.~\ref{parameter}, $\Delta(r)=0$ admits  three  positive roots viz $r_1$, $r_2$, $r_3$ and we address it as LMRBH with additional interior horizon (cf. Fig. \ref{hor1}). 
\item \textbf{NH} (peach colour region in Fig.~\ref{parameter}): $\Delta(r)=0$ has only one root (cf. Fig. \ref{hor1}) which is negative. This region is not of physical interest being ruled out by EHT observations of Sgr A* and is referred to as no-horizon (NH) spacetime.
\item \textbf{EBH} (Red  line): This represents an extremal LMRBH with degenerate horizons ($r_2=r_1$), where the outer two horizons merge. $\Delta(r)=0$ also admits one more negative root $r_3$ (cf. Fig. \ref{pen5}). At this line, there is a  transition from BH-II - NH or vice-versa.
\item \textbf{Dashed Red line}: This is an extreme case of BH-III where, like EBH, the outer two horizons merge. But, unlike EBH, the third root $r_3$ of $\Delta(r)=0$ is positive (cf. Fig. \ref{hor2}). At this line, the transition occurs from BH-III - BH-I or vice-versa.
\item \textbf{Blue line}: It represents a black hole with two degenerate horizons as a limiting case of BH-III where the inner two horizons merge. $\Delta(r)=0$ admits two equal positive roots $r_2=r_3$ and one larger root $r_1$ (cf. Fig. \ref{pen5}). At this line, the transition takes place from BH-III - BH-I or vice-versa.
\item \textbf{Green  line}: It represents a black hole with two horizons, located at positive values of $r$, and a null throat surface $r=0$ (cf. Fig. \ref{hor2}). At this line, the transition happens from BH-III - BH-II or vice-versa.
\item \textbf{Dashed Green line}: Like, green line, the surface $r=0$ is null but, unlike it, $\Delta(r)=0$ has no other roots than $r_1=0$ (cf. Fig. \ref{hor1}). At this line, the transition happens from  BH-I - NH or vice-versa.
\item \textbf{Red Dot}: The transition lines red and green intersect at this point. It represents an extremal black hole with a null throat. Here, $\Delta(r)=0$ admits three roots, $r_3=0$ and $r_1=r_2 \approx 0.763M$ for ($a, l$) $\approx$ ($1.095M, 0.601M$) (cf. Fig. \ref{hor2}).
\item \textbf{Black Dot}: At this point, transition lines red and blue merge. It represents an extremal black hole with three degenerate horizons. Here, $\Delta(r)=0$ admits three equal roots, $r_1=r_2=r_3=0.5M$  for ($a, l$) $\approx$ ($1.139M, 0.707M$) and it is referred as \textit{ultra extremal black hole}. 
\end{itemize}

\section{Structure of rotating spacetime: Penrose diagrams}\label{sec4}
Penrose diagrams are the way to present the causal structure of spacetime on 2-D finite sheet of paper. In the diagrams the light cone structure remains exactly same as in the Minkowski spacetime i.e. the null rays run at $45\degree$ with respect to the space and time axis. So, from the diagrams we can easily get the information what spacetime points are causally connected to what other spacetime points. We have drawn the diagrams for the LMRBH spacetime, corresponding to all the different regions in the parameter space (Fig. \ref{parameter}), along the symmetric $z$-axis ($\theta=0$).

Penrose diagrams for the region BH-I are same as for its non-rotating counterpart given in \cite{Peltola:2008pa,Peltola:2009jm}. Like the non-rotating version, it has two exterior regions (I and I$'$), two black hole and white hole interior regions ($r=r_1$ to $r=0$), and two quantum corrected interior regions ($r=0$ to $r=-\infty$) and the surface $r=0$, lying in the regions II and II$'$, is spacelike and hidden behind the event horizon. Penrose diagrams for regions NH and dashed green line, $l>0.601M$ are similar as for BH-I except the fact that for region NH, the surface $r=0$ lies in the regions I and I$'$ of Penrose diagram for BH-I and is time-like while for the dashed green line, $l>0.601M$ the surface $r=0$ is null.

Penrose diagrams for region BH-II look similar to that of Kerr black hole except for the extra horizon at $r=r_3$. In the region beyond $r=r_3$, $t$ and $r$ interchange their roles, $r$ behaving as the time-coordinate while $t$ behaving as a space-coordinate. Like Kerr, one has event horizon at $r=r_1$ and a Cauchy horizon at $r=r_2$, but the classical ring singularity of the Kerr black hole is replaced by the timelike transition surface $r=0$ inside the event horizon.  Also, in contrast to Kerr black hole, Penrose diagrams for region BH-II have quantum corrected interior regions ($r_3$ to $r=-\infty$), and the diagrams continue ad infinitum in the vertical as well the horizontal direction.  Penrose diagrams for regions BH-III and green line, $l<0.601M$ are similar as for BH-II except for the fact that for region BH-III, the surface $r=0$ lies in the regions IV and IV$'$ of the Penrose diagram for BH-II and is spacelike, while for the green line, $l<0.601M$ the surface $r=0$ is null.

Penrose diagrams for the region EBH, $l<0.601M$ are similar to the Kerr extremal black hole except for the extra horizon at $r=r_3$. The surface $r=0$, lying in the regions II and II$'$ of the diagram, is timelike. Like Kerr extremal black hole, the diagram for EBH continues ad infinitum in the vertical direction. Penrose diagrams for the region represented by dashed red line, $l>0.601M$ and red dot, $a=1.095M, l=0.601M$ are similar as for EBH, $l<0.601M$ except the fact that for dashed red line, $l>0.601M$, the surface $r=0$ lies in the regions III and III$'$ of Penrose diagram for EBH, $l<0.601M$ and is spacelike while for red dot, $a=1.095M, l=0.601M$ the surface $r=0$ is null.

Penrose diagrams for the region represented by the blue line in the parameter space look very different from the Penrose diagrams for the Kerr extremal black hole or the Kerr black hole having two distinct horizons. Here, the inner two horizons merge at positive $r=r_2=r_3$ with $t$ and $r$ interchanging their roles beyond it, such that the $r=0$ surface is spacelike. The diagram continues ad infinitum in the horizontal direction.

Penrose diagram for the black dot ($a, l$) $\approx$ ($1.139M, 0.707M$), which represents an \textit{ultra extremal} black hole, is exactly same as for the region BH-I.

\subsection{Embedding diagrams}
We embed the equatorial plane of the LQG motivated rotating spacetime in 3-D Euclidean space. At $t=const$ and equatorial plane ($\theta=\pi/2$), the metric (\ref{metric3}) reduces to the form
\begin{eqnarray}\label{embd1}
dl^2 &=&  \left[\frac{\rho^2}{\Delta} dr^2+ \frac{\mathcal{A}}{\rho^2} d\phi^2\right]_{\theta=\pi/2}.
\end{eqnarray}
The metric of 3-D Euclidean space in cylindrical coordinates can be written as
\begin{eqnarray}
dL^2 &=& dZ^2+ dR^2+ R^2d\phi^2.
\end{eqnarray}
The embedded surface will be axially symmetric and thus can be described by the function $Z=Z(R)$. For convenience we parametrize  the embedding surface as $Z(r)$, $R(r)$ where $r$ is the radial coordinate of the metric (\ref{embd1}). The metric induced on the embedding surface from the Euclidean space is
\begin{eqnarray}\label{embd2}
dl^2 &=& \left[\left(\frac{dZ}{dr}\right)^2+ \left(\frac{dR}{dr}\right)^2\right]dr^2+ R^2d\phi^2.
\end{eqnarray}
We require that the metric (\ref{embd2}) should be equal to metric (\ref{embd1}). So, comparing them we obtain

\begin{eqnarray}\label{scalar1}
R(r) &=& \sqrt{\frac{A}{\rho^2}}|_{\theta=\pi/2},\\
\left(\frac{dZ}{dr}\right)^2+ \left(\frac{dR}{dr}\right)^2 &=& \frac{\rho^2}{\Delta}|_{\theta=\pi/2},
\end{eqnarray}
which gives

\begin{eqnarray}\label{scalar2}
Z(r) &=& \int \,dr \sqrt{\frac{\rho^2}{\Delta}-\left(\frac{dR}{dr}\right)^2}|_{\theta=\pi/2} ~~ + ~const
\end{eqnarray}
\begin{figure*} 
	\begin{centering}
		\begin{tabular}{p{6cm} p{6cm} p{6cm}}
		 \includegraphics[scale=0.55]{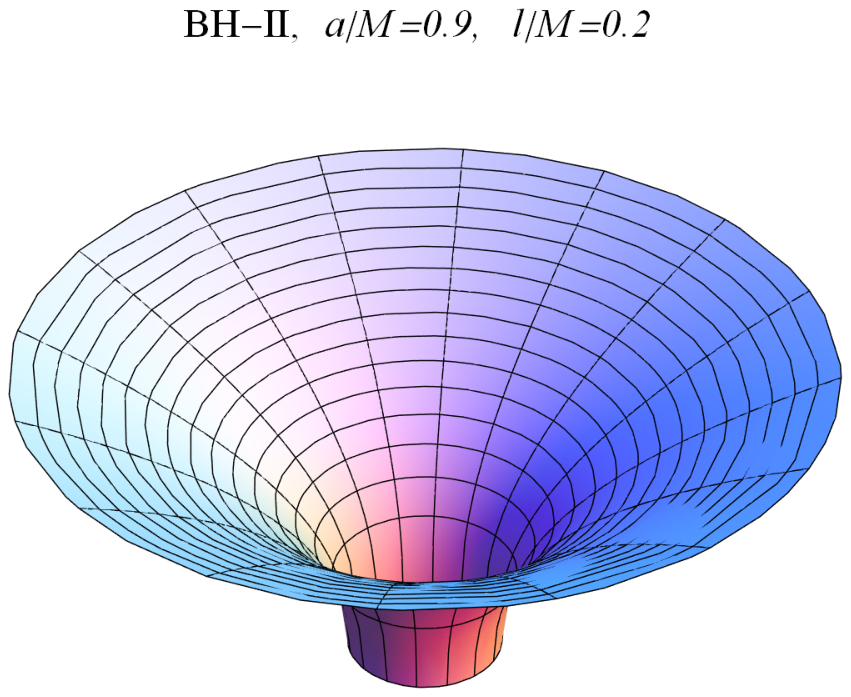}&
		 \includegraphics[scale=0.55]{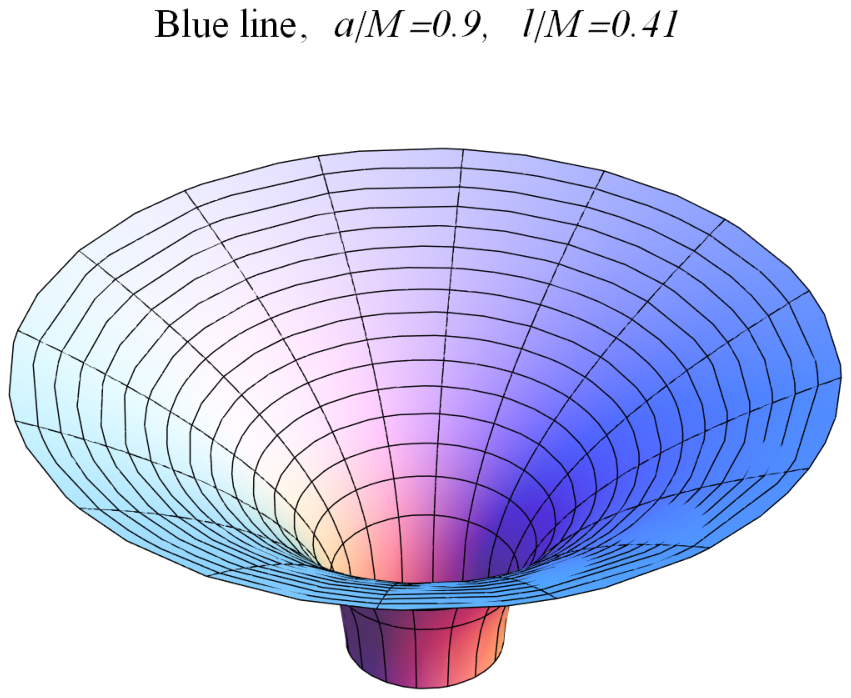}&
		 \includegraphics[scale=0.55]{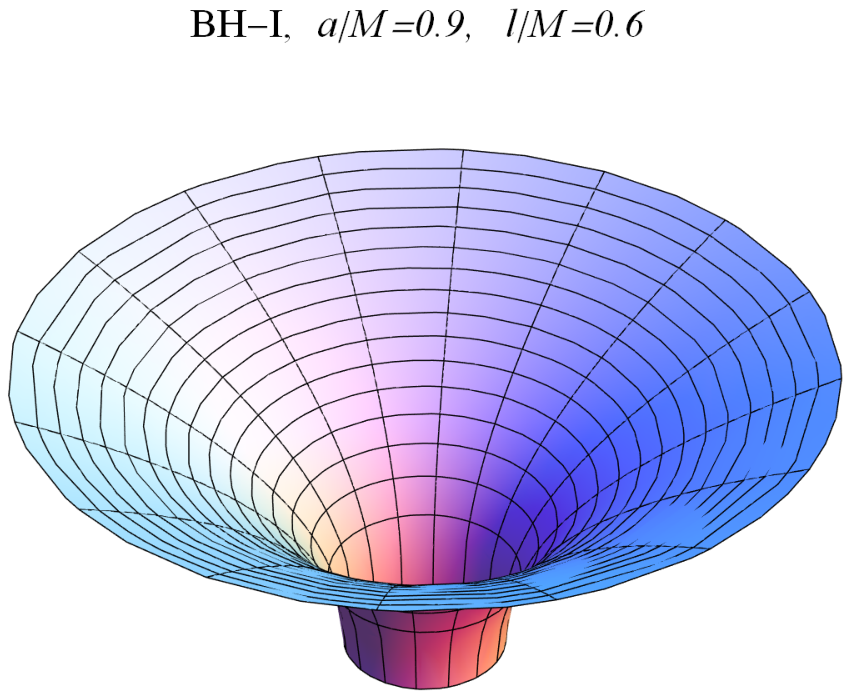}\\
		 \includegraphics[scale=0.55]{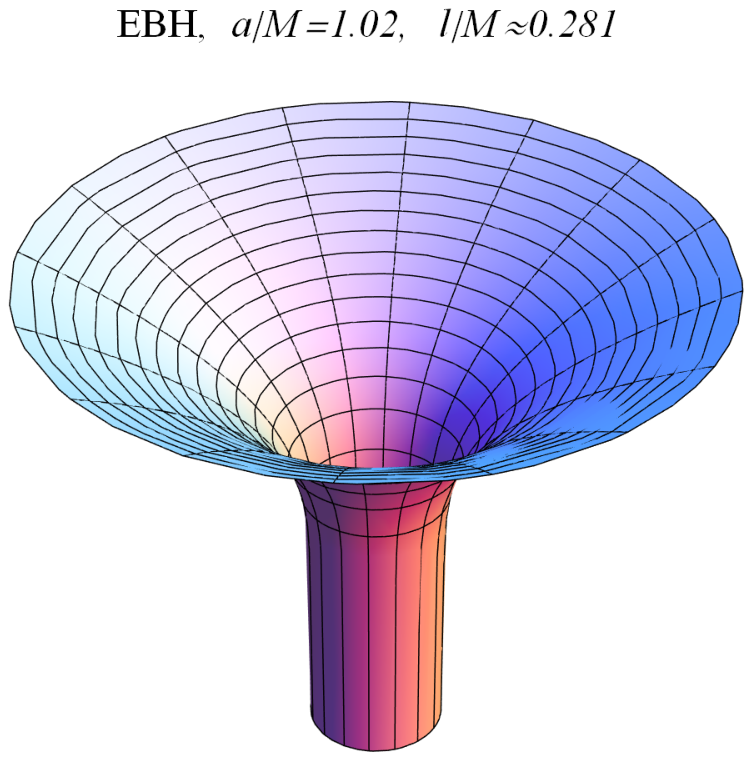}&
		 \includegraphics[scale=0.55]{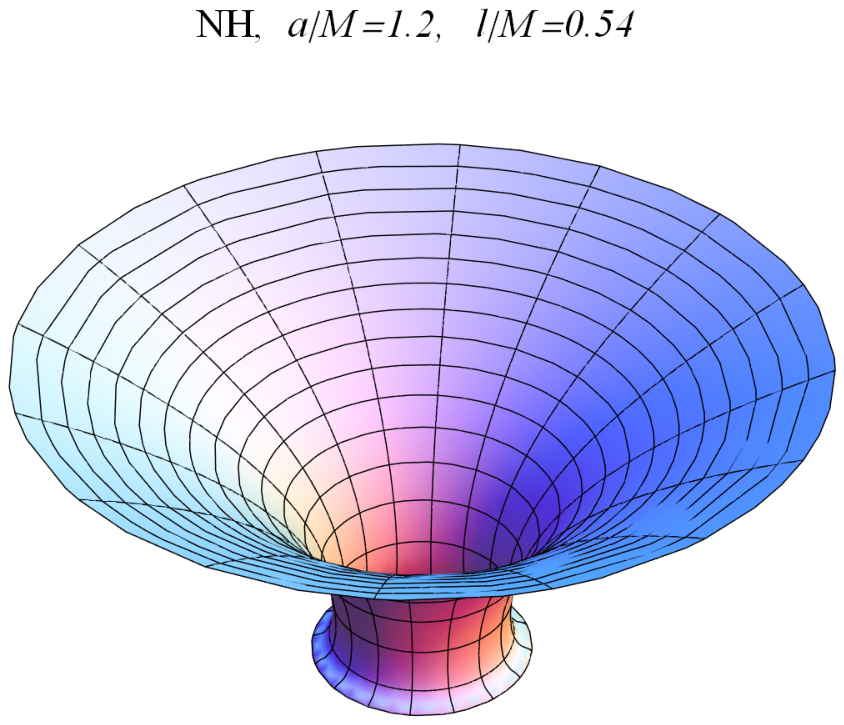}&
		 \includegraphics[scale=0.55]{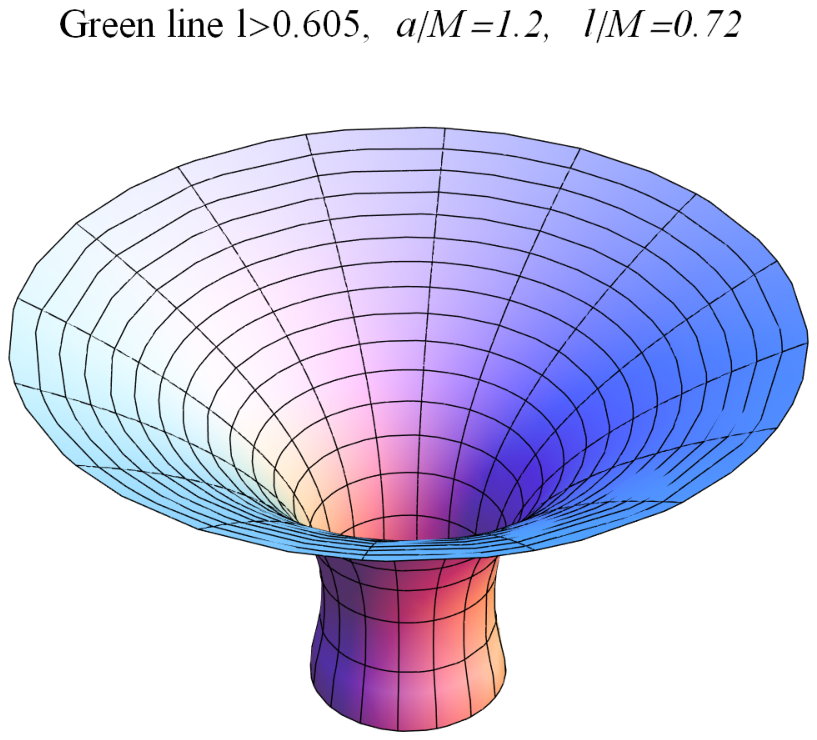}
			\end{tabular}
	\end{centering}
	\caption{Embedding diagrams, from event horizon to some finite radius $r$, in the equatorial plane ($\theta =\pi/2$) for selected regions of the parameter space ($M,a,l$). They are embeddable up to $r\to\infty$.  }\label{diagrams}
\end{figure*}
which can be solved numerically and, along with $R(r)$, gives the shape of the embedding surface provided ($\rho^2/\Delta-(dR/dr)^2)|_{\theta=\pi/2} \geq 0$. The embedding diagrams for the $\theta =\pi/2$ plane, corresponding to different regions in Fig. \ref{parameter}, are depicted in Fig. \ref{diagrams}. We can see from Eq. (\ref{scalar2}) that the $\theta =\pi/2$ plane is not embeddable when $\Delta$ is negative as the quantity under the square root becomes negative. Therefore, the upper three diagrams in Fig. \ref{diagrams} are the embedding diagrams for the region outside the event horizon. The boundary at the throat corresponds to the event horizon. There is no upper limit on $r$, and the diagrams are embeddable up to $r=\infty$. When we approach the EBH region, the throat in the embedding diagram gets longer and longer, and for the EBH region, it becomes infinitely long. For the embedding diagram EBH, we have chosen the values $a=1.02M$ and $l = 0.281M \approx l_c$, which are close to but not exactly in the EBH region and lie in the region BH-II. For ($a=1.2M$, $l=0.54M$) and ($a=1.2M$, $l=0.72M$), which respectively belong to the NH region and dashed green line, $l>0.601M$, the $\theta =\pi/2$ plane is embeddable for all positive values of $r$. The boundaries at the throats of these two diagrams correspond to the $r=0$ surface.

\section{Discussion and Conclusion}\label{sec5}
The Kerr metric well describes astrophysical black holes. It is the only stationary, vacuum axisymmetric metric that satisfies Einstein field equations and does not have pathologies outside the event horizon. The Event Horizon Telescope (EHT) observation unveiled the  images of supermassive black holes Sgr A* and M87*; \citep{EventHorizonTelescope:2019dse,EventHorizonTelescope:2019ggy,EventHorizonTelescope:2022xnr,EventHorizonTelescope:2022xqj}
The shadow size is within $~10\%$ of the Kerr predictions, providing us with another tool to investigate the nature of strong-field gravity. One can put constraints on potential deviations from the Kerr, i.e., Kerr-like black holes arising in modified gravity from  EHT observations. Hence, it could test fundamental quantum gravity theories, such as LQG,  as the quantum effects cannot be ignored in the strong-field regime. A rotating black hole plays a vital role in testing the quantum effects of gravity in this regime. The dearth of rotating black hole models in LQG substantially hampers the advancement of testing LQG from observations.  

Here,  we have handled this problem, we have used revised Newman-Janis algorithm - a viable solution-generating method to
generate a rotating solution from a nonrotating 
seed metric (\ref{metric2}). The resulting LMRBH spacetime (\ref{metric3}) possesses Kerr-like mathematical and several other interesting properties and exhibits rich spacetime structure. Unlike Kerr black hole, the $a=M$ case does not yield an extremal black hole in the LMRBH spacetime. We quantified this by explicitly solving the $\Delta(r)=0$ for real roots and demonstrating that the black hole spin yields around $a \approx 1.082M,\, 1.09M$, respectively, for $l = 0.56M,\, 0.6M$. Interestingly, the $\Delta(r)=0$ of LMRBH spacetime can admit up to three real roots, viz., $r_3,\, r_2,\; r_1$ with the possibility of the smallest root $r_3$ being negative. 
When all roots are positive, $r_2,\; r_1$, like Kerr black hole, are Cauchy and event horizons, and $r_3$ is an additional horizon inside Cauchy horizon.  The case of two positive roots is what we call the \textit{generic black hole}. As usual, $r_2=r_1$ gives an extremal black hole (cf. Fig. \ref{parameter} - red line). For dashed red line also black hole admits $r_2=r_1$ but on one side of the line the black hole admits three horizons while on the other side has one horizon (cf. Fig. \ref{parameter}). At the blue line  (cf. Fig. \ref{parameter}), in contrast to the red line, inner two horizons merge with $r_3=r_2$ and it is a transition line between regions BH-III and BH-I. The red and green lines intersect at red dot where again $r_2=r_1$, but $r_3=0$. Also the black dot in Fig. \ref{parameter} refers to the \textit{ultra extremal black hole} where $r_3 = r_2 = r_1$. On the green line the $\Delta(r)=0$ admits two positive roots, viz., $r_2,r_1$ and $r_3=0$, while on dashed green line $r_1=0$ is the only root. As another possibility,   $\Delta(r)=0$ has only one positive root, which is like spherical counterpart of LMRBH with single horizon. However, the region of spacetime where $\Delta(r)$ has negative root is not relevant for our discussion as that region is ruled out by EHT observations \cite{Shafqat:2022}.  

As our result exhibits, the LMRBH has exciting properties, e.g., a transition surface substitutes the classical ring singularity in  Kerr black hole at ($r=0, ~\theta =\pi/2$) and therby it provides sigularity resolution of Kerr black hole. 
Thus, we can expect that LMRBH captures some description of LQG. 

\section{Acknowledgments}  S.U.I and S.G.G. are supported by SERB-DST through project No.~CRG/2021/005771. J.K. would like to thank CSIR for providing SRF.
\bibliography{LQGM}

\begin{thebibliography}{69}%
\makeatletter
\providecommand \@ifxundefined [1]{%
 \@ifx{#1\undefined}
}%
\providecommand \@ifnum [1]{%
 \ifnum #1\expandafter \@firstoftwo
 \else \expandafter \@secondoftwo
 \fi
}%
\providecommand \@ifx [1]{%
 \ifx #1\expandafter \@firstoftwo
 \else \expandafter \@secondoftwo
 \fi
}%
\providecommand \natexlab [1]{#1}%
\providecommand \enquote  [1]{``#1''}%
\providecommand \bibnamefont  [1]{#1}%
\providecommand \bibfnamefont [1]{#1}%
\providecommand \citenamefont [1]{#1}%
\providecommand \href@noop [0]{\@secondoftwo}%
\providecommand \href [0]{\begingroup \@sanitize@url \@href}%
\providecommand \@href[1]{\@@startlink{#1}\@@href}%
\providecommand \@@href[1]{\endgroup#1\@@endlink}%
\providecommand \@sanitize@url [0]{\catcode `\\12\catcode `\$12\catcode
  `\&12\catcode `\#12\catcode `\^12\catcode `\_12\catcode `\%12\relax}%
\providecommand \@@startlink[1]{}%
\providecommand \@@endlink[0]{}%
\providecommand \url  [0]{\begingroup\@sanitize@url \@url }%
\providecommand \@url [1]{\endgroup\@href {#1}{\urlprefix }}%
\providecommand \urlprefix  [0]{URL }%
\providecommand \Eprint [0]{\href }%
\providecommand \doibase [0]{http://dx.doi.org/}%
\providecommand \selectlanguage [0]{\@gobble}%
\providecommand \bibinfo  [0]{\@secondoftwo}%
\providecommand \bibfield  [0]{\@secondoftwo}%
\providecommand \translation [1]{[#1]}%
\providecommand \BibitemOpen [0]{}%
\providecommand \bibitemStop [0]{}%
\providecommand \bibitemNoStop [0]{.\EOS\space}%
\providecommand \EOS [0]{\spacefactor3000\relax}%
\providecommand \BibitemShut  [1]{\csname bibitem#1\endcsname}%
\let\auto@bib@innerbib\@empty
\bibitem [{\citenamefont {Hawking}\ and\ \citenamefont
  {Penrose}(1970)}]{Hawking:1970zqf}%
  \BibitemOpen
  \bibfield  {author} {\bibinfo {author} {\bibfnamefont {S.~W.}\ \bibnamefont
  {Hawking}}\ and\ \bibinfo {author} {\bibfnamefont {R.}~\bibnamefont
  {Penrose}},\ }\href {\doibase 10.1098/rspa.1970.0021} {\bibfield  {journal}
  {\bibinfo  {journal} {Proc. Roy. Soc. Lond. A}\ }\textbf {\bibinfo {volume}
  {314}},\ \bibinfo {pages} {529--548} (\bibinfo {year} {1970})}\BibitemShut
  {NoStop}%
\bibitem [{\citenamefont {Hawking}\ and\ \citenamefont
  {Ellis}(1973)}]{Hawking:1973}%
  \BibitemOpen
  \bibfield  {author} {\bibinfo {author} {\bibfnamefont {S.}~\bibnamefont
  {Hawking}}\ and\ \bibinfo {author} {\bibfnamefont {G.}~\bibnamefont
  {Ellis}},\ }\href {\doibase 10.1103/PhysRevD.99.124012} {\bibfield  {journal}
  {\bibinfo  {journal} {Cambridge Univ. Press}\ }\textbf {\bibinfo {volume}
  {99}},\ \bibinfo {pages} {124012} (\bibinfo {year} {1973})}\BibitemShut
  {NoStop}%
\bibitem [{\citenamefont {Wheeler}(1963)}]{Wheeler:1964}%
  \BibitemOpen
  \bibfield  {author} {\bibinfo {author} {\bibfnamefont {J.~A.}\ \bibnamefont
  {Wheeler}},\ }\href@noop {} {\emph {\bibinfo {title} {{Relativity, Groups and
  Topology}}}}\ (\bibinfo  {publisher} {Gordon and Breach},\ \bibinfo {address}
  {New York},\ \bibinfo {year} {1963})\BibitemShut {NoStop}%
\bibitem [{\citenamefont {Ashtekar}\ \emph {et~al.}(2006)\citenamefont
  {Ashtekar}, \citenamefont {Pawlowski},\ and\ \citenamefont
  {Singh}}]{Ashtekar:2006wn}%
  \BibitemOpen
  \bibfield  {author} {\bibinfo {author} {\bibfnamefont {A.}~\bibnamefont
  {Ashtekar}}, \bibinfo {author} {\bibfnamefont {T.}~\bibnamefont {Pawlowski}},
  \ and\ \bibinfo {author} {\bibfnamefont {P.}~\bibnamefont {Singh}},\ }\href
  {\doibase 10.1103/PhysRevD.74.084003} {\bibfield  {journal} {\bibinfo
  {journal} {Phys. Rev. D}\ }\textbf {\bibinfo {volume} {74}},\ \bibinfo
  {pages} {084003} (\bibinfo {year} {2006})},\ \Eprint
  {http://arxiv.org/abs/gr-qc/0607039} {arXiv:gr-qc/0607039} \BibitemShut
  {NoStop}%
\bibitem [{\citenamefont {Ashtekar}\ \emph {et~al.}(2007)\citenamefont
  {Ashtekar}, \citenamefont {Pawlowski}, \citenamefont {Singh},\ and\
  \citenamefont {Vandersloot}}]{Ashtekar:2006es}%
  \BibitemOpen
  \bibfield  {author} {\bibinfo {author} {\bibfnamefont {A.}~\bibnamefont
  {Ashtekar}}, \bibinfo {author} {\bibfnamefont {T.}~\bibnamefont {Pawlowski}},
  \bibinfo {author} {\bibfnamefont {P.}~\bibnamefont {Singh}}, \ and\ \bibinfo
  {author} {\bibfnamefont {K.}~\bibnamefont {Vandersloot}},\ }\href {\doibase
  10.1103/PhysRevD.75.024035} {\bibfield  {journal} {\bibinfo  {journal} {Phys.
  Rev. D}\ }\textbf {\bibinfo {volume} {75}},\ \bibinfo {pages} {024035}
  (\bibinfo {year} {2007})},\ \Eprint {http://arxiv.org/abs/gr-qc/0612104}
  {arXiv:gr-qc/0612104} \BibitemShut {NoStop}%
\bibitem [{\citenamefont {Vandersloot}(2007)}]{Vandersloot:2006ws}%
  \BibitemOpen
  \bibfield  {author} {\bibinfo {author} {\bibfnamefont {K.}~\bibnamefont
  {Vandersloot}},\ }\href {\doibase 10.1103/PhysRevD.75.023523} {\bibfield
  {journal} {\bibinfo  {journal} {Phys. Rev. D}\ }\textbf {\bibinfo {volume}
  {75}},\ \bibinfo {pages} {023523} (\bibinfo {year} {2007})},\ \Eprint
  {http://arxiv.org/abs/gr-qc/0612070} {arXiv:gr-qc/0612070} \BibitemShut
  {NoStop}%
\bibitem [{\citenamefont {Modesto}(2006{\natexlab{a}})}]{Modesto:2004wm}%
  \BibitemOpen
  \bibfield  {author} {\bibinfo {author} {\bibfnamefont {L.}~\bibnamefont
  {Modesto}},\ }\href {\doibase 10.1007/s10773-006-9188-y} {\bibfield
  {journal} {\bibinfo  {journal} {Int. J. Theor. Phys.}\ }\textbf {\bibinfo
  {volume} {45}},\ \bibinfo {pages} {2235--2246} (\bibinfo {year}
  {2006}{\natexlab{a}})},\ \Eprint {http://arxiv.org/abs/gr-qc/0411032}
  {arXiv:gr-qc/0411032} \BibitemShut {NoStop}%
\bibitem [{\citenamefont {Ashtekar}\ and\ \citenamefont
  {Bojowald}(2006)}]{Ashtekar:2005qt}%
  \BibitemOpen
  \bibfield  {author} {\bibinfo {author} {\bibfnamefont {A.}~\bibnamefont
  {Ashtekar}}\ and\ \bibinfo {author} {\bibfnamefont {M.}~\bibnamefont
  {Bojowald}},\ }\href {\doibase 10.1088/0264-9381/23/2/008} {\bibfield
  {journal} {\bibinfo  {journal} {Class. Quant. Grav.}\ }\textbf {\bibinfo
  {volume} {23}},\ \bibinfo {pages} {391--411} (\bibinfo {year} {2006})},\
  \Eprint {http://arxiv.org/abs/gr-qc/0509075} {arXiv:gr-qc/0509075}
  \BibitemShut {NoStop}%
\bibitem [{\citenamefont {Modesto}(2006{\natexlab{b}})}]{Modesto:2005zm}%
  \BibitemOpen
  \bibfield  {author} {\bibinfo {author} {\bibfnamefont {L.}~\bibnamefont
  {Modesto}},\ }\href {\doibase 10.1088/0264-9381/23/18/006} {\bibfield
  {journal} {\bibinfo  {journal} {Class. Quant. Grav.}\ }\textbf {\bibinfo
  {volume} {23}},\ \bibinfo {pages} {5587--5602} (\bibinfo {year}
  {2006}{\natexlab{b}})},\ \Eprint {http://arxiv.org/abs/gr-qc/0509078}
  {arXiv:gr-qc/0509078} \BibitemShut {NoStop}%
\bibitem [{\citenamefont {Bojowald}(2020)}]{Bojowald:2020dkb}%
  \BibitemOpen
  \bibfield  {author} {\bibinfo {author} {\bibfnamefont {M.}~\bibnamefont
  {Bojowald}},\ }\href {\doibase 10.3390/universe6080125} {\bibfield  {journal}
  {\bibinfo  {journal} {Universe}\ }\textbf {\bibinfo {volume} {6}},\ \bibinfo
  {pages} {125} (\bibinfo {year} {2020})},\ \Eprint
  {http://arxiv.org/abs/2009.13565} {arXiv:2009.13565 [gr-qc]} \BibitemShut
  {NoStop}%
\bibitem [{\citenamefont {Gambini}\ and\ \citenamefont
  {Pullin}(2013)}]{Gambini:2013ooa}%
  \BibitemOpen
  \bibfield  {author} {\bibinfo {author} {\bibfnamefont {R.}~\bibnamefont
  {Gambini}}\ and\ \bibinfo {author} {\bibfnamefont {J.}~\bibnamefont
  {Pullin}},\ }\href {\doibase 10.1103/PhysRevLett.110.211301} {\bibfield
  {journal} {\bibinfo  {journal} {Phys. Rev. Lett.}\ }\textbf {\bibinfo
  {volume} {110}},\ \bibinfo {pages} {211301} (\bibinfo {year} {2013})},\
  \Eprint {http://arxiv.org/abs/1302.5265} {arXiv:1302.5265 [gr-qc]}
  \BibitemShut {NoStop}%
\bibitem [{\citenamefont {Corichi}\ and\ \citenamefont
  {Singh}(2016)}]{Corichi:2015xia}%
  \BibitemOpen
  \bibfield  {author} {\bibinfo {author} {\bibfnamefont {A.}~\bibnamefont
  {Corichi}}\ and\ \bibinfo {author} {\bibfnamefont {P.}~\bibnamefont
  {Singh}},\ }\href {\doibase 10.1088/0264-9381/33/5/055006} {\bibfield
  {journal} {\bibinfo  {journal} {Class. Quant. Grav.}\ }\textbf {\bibinfo
  {volume} {33}},\ \bibinfo {pages} {055006} (\bibinfo {year} {2016})},\
  \Eprint {http://arxiv.org/abs/1506.08015} {arXiv:1506.08015 [gr-qc]}
  \BibitemShut {NoStop}%
\bibitem [{\citenamefont {Olmedo}\ \emph {et~al.}(2017)\citenamefont {Olmedo},
  \citenamefont {Saini},\ and\ \citenamefont {Singh}}]{Olmedo:2017lvt}%
  \BibitemOpen
  \bibfield  {author} {\bibinfo {author} {\bibfnamefont {J.}~\bibnamefont
  {Olmedo}}, \bibinfo {author} {\bibfnamefont {S.}~\bibnamefont {Saini}}, \
  and\ \bibinfo {author} {\bibfnamefont {P.}~\bibnamefont {Singh}},\ }\href
  {\doibase 10.1088/1361-6382/aa8da8} {\bibfield  {journal} {\bibinfo
  {journal} {Class. Quant. Grav.}\ }\textbf {\bibinfo {volume} {34}},\ \bibinfo
  {pages} {225011} (\bibinfo {year} {2017})},\ \Eprint
  {http://arxiv.org/abs/1707.07333} {arXiv:1707.07333 [gr-qc]} \BibitemShut
  {NoStop}%
\bibitem [{\citenamefont {Ashtekar}\ \emph
  {et~al.}(2018{\natexlab{a}})\citenamefont {Ashtekar}, \citenamefont
  {Olmedo},\ and\ \citenamefont {Singh}}]{Ashtekar:2018lag}%
  \BibitemOpen
  \bibfield  {author} {\bibinfo {author} {\bibfnamefont {A.}~\bibnamefont
  {Ashtekar}}, \bibinfo {author} {\bibfnamefont {J.}~\bibnamefont {Olmedo}}, \
  and\ \bibinfo {author} {\bibfnamefont {P.}~\bibnamefont {Singh}},\ }\href
  {\doibase 10.1103/PhysRevLett.121.241301} {\bibfield  {journal} {\bibinfo
  {journal} {Phys. Rev. Lett.}\ }\textbf {\bibinfo {volume} {121}},\ \bibinfo
  {pages} {241301} (\bibinfo {year} {2018}{\natexlab{a}})},\ \Eprint
  {http://arxiv.org/abs/1806.00648} {arXiv:1806.00648 [gr-qc]} \BibitemShut
  {NoStop}%
\bibitem [{\citenamefont {Ashtekar}\ \emph
  {et~al.}(2018{\natexlab{b}})\citenamefont {Ashtekar}, \citenamefont
  {Olmedo},\ and\ \citenamefont {Singh}}]{Ashtekar:2018cay}%
  \BibitemOpen
  \bibfield  {author} {\bibinfo {author} {\bibfnamefont {A.}~\bibnamefont
  {Ashtekar}}, \bibinfo {author} {\bibfnamefont {J.}~\bibnamefont {Olmedo}}, \
  and\ \bibinfo {author} {\bibfnamefont {P.}~\bibnamefont {Singh}},\ }\href
  {\doibase 10.1103/PhysRevD.98.126003} {\bibfield  {journal} {\bibinfo
  {journal} {Phys. Rev. D}\ }\textbf {\bibinfo {volume} {98}},\ \bibinfo
  {pages} {126003} (\bibinfo {year} {2018}{\natexlab{b}})},\ \Eprint
  {http://arxiv.org/abs/1806.02406} {arXiv:1806.02406 [gr-qc]} \BibitemShut
  {NoStop}%
\bibitem [{\citenamefont {Bodendorfer}\ \emph
  {et~al.}(2019{\natexlab{a}})\citenamefont {Bodendorfer}, \citenamefont
  {Mele},\ and\ \citenamefont {M\"unch}}]{Bodendorfer:2019xbp}%
  \BibitemOpen
  \bibfield  {author} {\bibinfo {author} {\bibfnamefont {N.}~\bibnamefont
  {Bodendorfer}}, \bibinfo {author} {\bibfnamefont {F.~M.}\ \bibnamefont
  {Mele}}, \ and\ \bibinfo {author} {\bibfnamefont {J.}~\bibnamefont
  {M\"unch}},\ }\href {\doibase 10.1088/1361-6382/ab32ba} {\bibfield  {journal}
  {\bibinfo  {journal} {Class. Quant. Grav.}\ }\textbf {\bibinfo {volume}
  {36}},\ \bibinfo {pages} {187001} (\bibinfo {year} {2019}{\natexlab{a}})},\
  \Eprint {http://arxiv.org/abs/1902.04032} {arXiv:1902.04032 [gr-qc]}
  \BibitemShut {NoStop}%
\bibitem [{\citenamefont {Bodendorfer}\ \emph
  {et~al.}(2019{\natexlab{b}})\citenamefont {Bodendorfer}, \citenamefont
  {Mele},\ and\ \citenamefont {M\"unch}}]{Bodendorfer:2019cyv}%
  \BibitemOpen
  \bibfield  {author} {\bibinfo {author} {\bibfnamefont {N.}~\bibnamefont
  {Bodendorfer}}, \bibinfo {author} {\bibfnamefont {F.~M.}\ \bibnamefont
  {Mele}}, \ and\ \bibinfo {author} {\bibfnamefont {J.}~\bibnamefont
  {M\"unch}},\ }\href {\doibase 10.1088/1361-6382/ab3f16} {\bibfield  {journal}
  {\bibinfo  {journal} {Class. Quant. Grav.}\ }\textbf {\bibinfo {volume}
  {36}},\ \bibinfo {pages} {195015} (\bibinfo {year} {2019}{\natexlab{b}})},\
  \Eprint {http://arxiv.org/abs/1902.04542} {arXiv:1902.04542 [gr-qc]}
  \BibitemShut {NoStop}%
\bibitem [{\citenamefont {Arruga}\ \emph {et~al.}(2020)\citenamefont {Arruga},
  \citenamefont {Ben~Achour},\ and\ \citenamefont {Noui}}]{Arruga:2019kyd}%
  \BibitemOpen
  \bibfield  {author} {\bibinfo {author} {\bibfnamefont {D.}~\bibnamefont
  {Arruga}}, \bibinfo {author} {\bibfnamefont {J.}~\bibnamefont {Ben~Achour}},
  \ and\ \bibinfo {author} {\bibfnamefont {K.}~\bibnamefont {Noui}},\ }\href
  {\doibase 10.3390/universe6030039} {\bibfield  {journal} {\bibinfo  {journal}
  {Universe}\ }\textbf {\bibinfo {volume} {6}},\ \bibinfo {pages} {39}
  (\bibinfo {year} {2020})},\ \Eprint {http://arxiv.org/abs/1912.02459}
  {arXiv:1912.02459 [gr-qc]} \BibitemShut {NoStop}%
\bibitem [{\citenamefont {Assanioussi}\ \emph {et~al.}(2020)\citenamefont
  {Assanioussi}, \citenamefont {Dapor},\ and\ \citenamefont
  {Liegener}}]{Assanioussi:2019twp}%
  \BibitemOpen
  \bibfield  {author} {\bibinfo {author} {\bibfnamefont {M.}~\bibnamefont
  {Assanioussi}}, \bibinfo {author} {\bibfnamefont {A.}~\bibnamefont {Dapor}},
  \ and\ \bibinfo {author} {\bibfnamefont {K.}~\bibnamefont {Liegener}},\
  }\href {\doibase 10.1103/PhysRevD.101.026002} {\bibfield  {journal} {\bibinfo
   {journal} {Phys. Rev. D}\ }\textbf {\bibinfo {volume} {101}},\ \bibinfo
  {pages} {026002} (\bibinfo {year} {2020})},\ \Eprint
  {http://arxiv.org/abs/1908.05756} {arXiv:1908.05756 [gr-qc]} \BibitemShut
  {NoStop}%
\bibitem [{\citenamefont {Ben~Achour}\ \emph {et~al.}(2020)\citenamefont
  {Ben~Achour}, \citenamefont {Brahma}, \citenamefont {Mukohyama},\ and\
  \citenamefont {Uzan}}]{BenAchour:2020gon}%
  \BibitemOpen
  \bibfield  {author} {\bibinfo {author} {\bibfnamefont {J.}~\bibnamefont
  {Ben~Achour}}, \bibinfo {author} {\bibfnamefont {S.}~\bibnamefont {Brahma}},
  \bibinfo {author} {\bibfnamefont {S.}~\bibnamefont {Mukohyama}}, \ and\
  \bibinfo {author} {\bibfnamefont {J.~P.}\ \bibnamefont {Uzan}},\ }\href
  {\doibase 10.1088/1475-7516/2020/09/020} {\bibfield  {journal} {\bibinfo
  {journal} {JCAP}\ }\textbf {\bibinfo {volume} {09}},\ \bibinfo {pages} {020}
  (\bibinfo {year} {2020})},\ \Eprint {http://arxiv.org/abs/2004.12977}
  {arXiv:2004.12977 [gr-qc]} \BibitemShut {NoStop}%
\bibitem [{\citenamefont {Gambini}\ \emph {et~al.}(2020)\citenamefont
  {Gambini}, \citenamefont {Olmedo},\ and\ \citenamefont
  {Pullin}}]{Gambini:2020nsf}%
  \BibitemOpen
  \bibfield  {author} {\bibinfo {author} {\bibfnamefont {R.}~\bibnamefont
  {Gambini}}, \bibinfo {author} {\bibfnamefont {J.}~\bibnamefont {Olmedo}}, \
  and\ \bibinfo {author} {\bibfnamefont {J.}~\bibnamefont {Pullin}},\ }\href
  {\doibase 10.1088/1361-6382/aba842} {\bibfield  {journal} {\bibinfo
  {journal} {Class. Quant. Grav.}\ }\textbf {\bibinfo {volume} {37}},\ \bibinfo
  {pages} {205012} (\bibinfo {year} {2020})},\ \Eprint
  {http://arxiv.org/abs/2006.01513} {arXiv:2006.01513 [gr-qc]} \BibitemShut
  {NoStop}%
\bibitem [{\citenamefont {Bodendorfer}\ \emph
  {et~al.}(2021{\natexlab{a}})\citenamefont {Bodendorfer}, \citenamefont
  {Mele},\ and\ \citenamefont {M\"unch}}]{Bodendorfer:2019nvy}%
  \BibitemOpen
  \bibfield  {author} {\bibinfo {author} {\bibfnamefont {N.}~\bibnamefont
  {Bodendorfer}}, \bibinfo {author} {\bibfnamefont {F.~M.}\ \bibnamefont
  {Mele}}, \ and\ \bibinfo {author} {\bibfnamefont {J.}~\bibnamefont
  {M\"unch}},\ }\href {\doibase 10.1016/j.physletb.2021.136390} {\bibfield
  {journal} {\bibinfo  {journal} {Phys. Lett. B}\ }\textbf {\bibinfo {volume}
  {819}},\ \bibinfo {pages} {136390} (\bibinfo {year} {2021}{\natexlab{a}})},\
  \Eprint {http://arxiv.org/abs/1911.12646} {arXiv:1911.12646 [gr-qc]}
  \BibitemShut {NoStop}%
\bibitem [{\citenamefont {Bodendorfer}\ \emph
  {et~al.}(2021{\natexlab{b}})\citenamefont {Bodendorfer}, \citenamefont
  {Mele},\ and\ \citenamefont {M\"unch}}]{Bodendorfer:2019jay}%
  \BibitemOpen
  \bibfield  {author} {\bibinfo {author} {\bibfnamefont {N.}~\bibnamefont
  {Bodendorfer}}, \bibinfo {author} {\bibfnamefont {F.~M.}\ \bibnamefont
  {Mele}}, \ and\ \bibinfo {author} {\bibfnamefont {J.}~\bibnamefont
  {M\"unch}},\ }\href {\doibase 10.1088/1361-6382/abe05d} {\bibfield  {journal}
  {\bibinfo  {journal} {Class. Quant. Grav.}\ }\textbf {\bibinfo {volume}
  {38}},\ \bibinfo {pages} {095002} (\bibinfo {year} {2021}{\natexlab{b}})},\
  \Eprint {http://arxiv.org/abs/1912.00774} {arXiv:1912.00774 [gr-qc]}
  \BibitemShut {NoStop}%
\bibitem [{\citenamefont {Blanchette}\ \emph {et~al.}(2021)\citenamefont
  {Blanchette}, \citenamefont {Das}, \citenamefont {Hergott},\ and\
  \citenamefont {Rastgoo}}]{Blanchette:2020kkk}%
  \BibitemOpen
  \bibfield  {author} {\bibinfo {author} {\bibfnamefont {K.}~\bibnamefont
  {Blanchette}}, \bibinfo {author} {\bibfnamefont {S.}~\bibnamefont {Das}},
  \bibinfo {author} {\bibfnamefont {S.}~\bibnamefont {Hergott}}, \ and\
  \bibinfo {author} {\bibfnamefont {S.}~\bibnamefont {Rastgoo}},\ }\href
  {\doibase 10.1103/PhysRevD.103.084038} {\bibfield  {journal} {\bibinfo
  {journal} {Phys. Rev. D}\ }\textbf {\bibinfo {volume} {103}},\ \bibinfo
  {pages} {084038} (\bibinfo {year} {2021})},\ \Eprint
  {http://arxiv.org/abs/2011.11815} {arXiv:2011.11815 [gr-qc]} \BibitemShut
  {NoStop}%
\bibitem [{\citenamefont {Assanioussi}\ and\ \citenamefont
  {Mickel}(2021)}]{Assanioussi:2020ezr}%
  \BibitemOpen
  \bibfield  {author} {\bibinfo {author} {\bibfnamefont {M.}~\bibnamefont
  {Assanioussi}}\ and\ \bibinfo {author} {\bibfnamefont {L.}~\bibnamefont
  {Mickel}},\ }\href {\doibase 10.1103/PhysRevD.103.124008} {\bibfield
  {journal} {\bibinfo  {journal} {Phys. Rev. D}\ }\textbf {\bibinfo {volume}
  {103}},\ \bibinfo {pages} {124008} (\bibinfo {year} {2021})},\ \Eprint
  {http://arxiv.org/abs/2012.06839} {arXiv:2012.06839 [gr-qc]} \BibitemShut
  {NoStop}%
\bibitem [{\citenamefont {Chen}(2022)}]{Chen:2022nix}%
  \BibitemOpen
  \bibfield  {author} {\bibinfo {author} {\bibfnamefont {C.-Y.}\ \bibnamefont
  {Chen}},\ }in\ \href@noop {} {\emph {\bibinfo {booktitle} {{Geometric
  Foundations of Gravity 2021}}}}\ (\bibinfo {year} {2022})\ \Eprint
  {http://arxiv.org/abs/2207.03797} {arXiv:2207.03797 [gr-qc]} \BibitemShut
  {NoStop}%
\bibitem [{\citenamefont {Bojowald}(2005)}]{Bojowald:2005epg}%
  \BibitemOpen
  \bibfield  {author} {\bibinfo {author} {\bibfnamefont {M.}~\bibnamefont
  {Bojowald}},\ }\href {\doibase 10.12942/lrr-2005-11} {\bibfield  {journal}
  {\bibinfo  {journal} {Living Rev. Rel.}\ }\textbf {\bibinfo {volume} {8}},\
  \bibinfo {pages} {11} (\bibinfo {year} {2005})},\ \Eprint
  {http://arxiv.org/abs/gr-qc/0601085} {arXiv:gr-qc/0601085} \BibitemShut
  {NoStop}%
\bibitem [{\citenamefont {Singh}(2009)}]{Singh:2009mz}%
  \BibitemOpen
  \bibfield  {author} {\bibinfo {author} {\bibfnamefont {P.}~\bibnamefont
  {Singh}},\ }\href {\doibase 10.1088/0264-9381/26/12/125005} {\bibfield
  {journal} {\bibinfo  {journal} {Class. Quant. Grav.}\ }\textbf {\bibinfo
  {volume} {26}},\ \bibinfo {pages} {125005} (\bibinfo {year} {2009})},\
  \Eprint {http://arxiv.org/abs/0901.2750} {arXiv:0901.2750 [gr-qc]}
  \BibitemShut {NoStop}%
\bibitem [{\citenamefont {Zhang}(2009)}]{Zhang:2007yu}%
  \BibitemOpen
  \bibfield  {author} {\bibinfo {author} {\bibfnamefont {X.}~\bibnamefont
  {Zhang}},\ }\href {\doibase 10.1140/epjc/s10052-009-0967-5} {\bibfield
  {journal} {\bibinfo  {journal} {Eur. Phys. J. C}\ }\textbf {\bibinfo {volume}
  {60}},\ \bibinfo {pages} {661--667} (\bibinfo {year} {2009})},\ \Eprint
  {http://arxiv.org/abs/0708.1408} {arXiv:0708.1408 [gr-qc]} \BibitemShut
  {NoStop}%
\bibitem [{\citenamefont {Kumar~Walia}(2022)}]{KumarWalia:2022ddq}%
  \BibitemOpen
  \bibfield  {author} {\bibinfo {author} {\bibfnamefont {R.}~\bibnamefont
  {Kumar~Walia}},\ }\href@noop {} {\  (\bibinfo {year} {2022})},\ \Eprint
  {http://arxiv.org/abs/2207.02106} {arXiv:2207.02106 [gr-qc]} \BibitemShut
  {NoStop}%
\bibitem [{\citenamefont {Ashtekar}\ \emph {et~al.}(2003)\citenamefont
  {Ashtekar}, \citenamefont {Fairhurst},\ and\ \citenamefont
  {Willis}}]{Ashtekar:2002sn}%
  \BibitemOpen
  \bibfield  {author} {\bibinfo {author} {\bibfnamefont {A.}~\bibnamefont
  {Ashtekar}}, \bibinfo {author} {\bibfnamefont {S.}~\bibnamefont {Fairhurst}},
  \ and\ \bibinfo {author} {\bibfnamefont {J.~L.}\ \bibnamefont {Willis}},\
  }\href {\doibase 10.1088/0264-9381/20/6/302} {\bibfield  {journal} {\bibinfo
  {journal} {Class. Quant. Grav.}\ }\textbf {\bibinfo {volume} {20}},\ \bibinfo
  {pages} {1031--1062} (\bibinfo {year} {2003})},\ \Eprint
  {http://arxiv.org/abs/gr-qc/0207106} {arXiv:gr-qc/0207106} \BibitemShut
  {NoStop}%
\bibitem [{\citenamefont {Boehmer}\ and\ \citenamefont
  {Vandersloot}(2007)}]{Boehmer:2007ket}%
  \BibitemOpen
  \bibfield  {author} {\bibinfo {author} {\bibfnamefont {C.~G.}\ \bibnamefont
  {Boehmer}}\ and\ \bibinfo {author} {\bibfnamefont {K.}~\bibnamefont
  {Vandersloot}},\ }\href {\doibase 10.1103/PhysRevD.76.104030} {\bibfield
  {journal} {\bibinfo  {journal} {Phys. Rev. D}\ }\textbf {\bibinfo {volume}
  {76}},\ \bibinfo {pages} {104030} (\bibinfo {year} {2007})},\ \Eprint
  {http://arxiv.org/abs/0709.2129} {arXiv:0709.2129 [gr-qc]} \BibitemShut
  {NoStop}%
\bibitem [{\citenamefont {Peltola}\ and\ \citenamefont
  {Kunstatter}(2009{\natexlab{a}})}]{Peltola:2008pa}%
  \BibitemOpen
  \bibfield  {author} {\bibinfo {author} {\bibfnamefont {A.}~\bibnamefont
  {Peltola}}\ and\ \bibinfo {author} {\bibfnamefont {G.}~\bibnamefont
  {Kunstatter}},\ }\href {\doibase 10.1103/PhysRevD.79.061501} {\bibfield
  {journal} {\bibinfo  {journal} {Phys. Rev. D}\ }\textbf {\bibinfo {volume}
  {79}},\ \bibinfo {pages} {061501} (\bibinfo {year} {2009}{\natexlab{a}})},\
  \Eprint {http://arxiv.org/abs/0811.3240} {arXiv:0811.3240 [gr-qc]}
  \BibitemShut {NoStop}%
\bibitem [{\citenamefont {Ashtekar}\ and\ \citenamefont
  {Lewandowski}(2004)}]{Ashtekar:2004eh}%
  \BibitemOpen
  \bibfield  {author} {\bibinfo {author} {\bibfnamefont {A.}~\bibnamefont
  {Ashtekar}}\ and\ \bibinfo {author} {\bibfnamefont {J.}~\bibnamefont
  {Lewandowski}},\ }\href {\doibase 10.1088/0264-9381/21/15/R01} {\bibfield
  {journal} {\bibinfo  {journal} {Class. Quant. Grav.}\ }\textbf {\bibinfo
  {volume} {21}},\ \bibinfo {pages} {R53} (\bibinfo {year} {2004})},\ \Eprint
  {http://arxiv.org/abs/gr-qc/0404018} {arXiv:gr-qc/0404018} \BibitemShut
  {NoStop}%
\bibitem [{\citenamefont {Peltola}\ and\ \citenamefont
  {Kunstatter}(2009{\natexlab{b}})}]{Peltola:2009jm}%
  \BibitemOpen
  \bibfield  {author} {\bibinfo {author} {\bibfnamefont {A.}~\bibnamefont
  {Peltola}}\ and\ \bibinfo {author} {\bibfnamefont {G.}~\bibnamefont
  {Kunstatter}},\ }\href {\doibase 10.1103/PhysRevD.80.044031} {\bibfield
  {journal} {\bibinfo  {journal} {Phys. Rev. D}\ }\textbf {\bibinfo {volume}
  {80}},\ \bibinfo {pages} {044031} (\bibinfo {year} {2009}{\natexlab{b}})},\
  \Eprint {http://arxiv.org/abs/0902.1746} {arXiv:0902.1746 [gr-qc]}
  \BibitemShut {NoStop}%
\bibitem [{\citenamefont {Modesto}(2010)}]{Modesto:2008im}%
  \BibitemOpen
  \bibfield  {author} {\bibinfo {author} {\bibfnamefont {L.}~\bibnamefont
  {Modesto}},\ }\href {\doibase 10.1007/s10773-010-0346-x} {\bibfield
  {journal} {\bibinfo  {journal} {Int. J. Theor. Phys.}\ }\textbf {\bibinfo
  {volume} {49}},\ \bibinfo {pages} {1649--1683} (\bibinfo {year} {2010})},\
  \Eprint {http://arxiv.org/abs/0811.2196} {arXiv:0811.2196 [gr-qc]}
  \BibitemShut {NoStop}%
\bibitem [{\citenamefont {Modesto}(2008)}]{Modesto:2006mx}%
  \BibitemOpen
  \bibfield  {author} {\bibinfo {author} {\bibfnamefont {L.}~\bibnamefont
  {Modesto}},\ }\href {\doibase 10.1155/2008/459290} {\bibfield  {journal}
  {\bibinfo  {journal} {Adv. High Energy Phys.}\ }\textbf {\bibinfo {volume}
  {2008}},\ \bibinfo {pages} {459290} (\bibinfo {year} {2008})},\ \Eprint
  {http://arxiv.org/abs/gr-qc/0611043} {arXiv:gr-qc/0611043} \BibitemShut
  {NoStop}%
\bibitem [{\citenamefont {Boehmer}\ and\ \citenamefont
  {Vandersloot}(2008)}]{Boehmer:2008fz}%
  \BibitemOpen
  \bibfield  {author} {\bibinfo {author} {\bibfnamefont {C.~G.}\ \bibnamefont
  {Boehmer}}\ and\ \bibinfo {author} {\bibfnamefont {K.}~\bibnamefont
  {Vandersloot}},\ }\href {\doibase 10.1103/PhysRevD.78.067501} {\bibfield
  {journal} {\bibinfo  {journal} {Phys. Rev. D}\ }\textbf {\bibinfo {volume}
  {78}},\ \bibinfo {pages} {067501} (\bibinfo {year} {2008})},\ \Eprint
  {http://arxiv.org/abs/0807.3042} {arXiv:0807.3042 [gr-qc]} \BibitemShut
  {NoStop}%
\bibitem [{\citenamefont {Campiglia}\ \emph {et~al.}(2008)\citenamefont
  {Campiglia}, \citenamefont {Gambini},\ and\ \citenamefont
  {Pullin}}]{Campiglia:2007pb}%
  \BibitemOpen
  \bibfield  {author} {\bibinfo {author} {\bibfnamefont {M.}~\bibnamefont
  {Campiglia}}, \bibinfo {author} {\bibfnamefont {R.}~\bibnamefont {Gambini}},
  \ and\ \bibinfo {author} {\bibfnamefont {J.}~\bibnamefont {Pullin}},\ }\href
  {\doibase 10.1063/1.2902798} {\bibfield  {journal} {\bibinfo  {journal} {AIP
  Conf. Proc.}\ }\textbf {\bibinfo {volume} {977}},\ \bibinfo {pages} {52--63}
  (\bibinfo {year} {2008})},\ \Eprint {http://arxiv.org/abs/0712.0817}
  {arXiv:0712.0817 [gr-qc]} \BibitemShut {NoStop}%
\bibitem [{\citenamefont {Gambini}\ and\ \citenamefont
  {Pullin}(2008)}]{Gambini:2008dy}%
  \BibitemOpen
  \bibfield  {author} {\bibinfo {author} {\bibfnamefont {R.}~\bibnamefont
  {Gambini}}\ and\ \bibinfo {author} {\bibfnamefont {J.}~\bibnamefont
  {Pullin}},\ }\href {\doibase 10.1103/PhysRevLett.101.161301} {\bibfield
  {journal} {\bibinfo  {journal} {Phys. Rev. Lett.}\ }\textbf {\bibinfo
  {volume} {101}},\ \bibinfo {pages} {161301} (\bibinfo {year} {2008})},\
  \Eprint {http://arxiv.org/abs/0805.1187} {arXiv:0805.1187 [gr-qc]}
  \BibitemShut {NoStop}%
\bibitem [{\citenamefont {Kerr}(1963)}]{Kerr:1963ud}%
  \BibitemOpen
  \bibfield  {author} {\bibinfo {author} {\bibfnamefont {R.~P.}\ \bibnamefont
  {Kerr}},\ }\href {\doibase 10.1103/PhysRevLett.11.237} {\bibfield  {journal}
  {\bibinfo  {journal} {Phys. Rev. Lett.}\ }\textbf {\bibinfo {volume} {11}},\
  \bibinfo {pages} {237--238} (\bibinfo {year} {1963})}\BibitemShut {NoStop}%
\bibitem [{\citenamefont {Brahma}\ \emph {et~al.}(2021)\citenamefont {Brahma},
  \citenamefont {Chen},\ and\ \citenamefont {Yeom}}]{Brahma:2020eos}%
  \BibitemOpen
  \bibfield  {author} {\bibinfo {author} {\bibfnamefont {S.}~\bibnamefont
  {Brahma}}, \bibinfo {author} {\bibfnamefont {C.-Y.}\ \bibnamefont {Chen}}, \
  and\ \bibinfo {author} {\bibfnamefont {D.-h.}\ \bibnamefont {Yeom}},\ }\href
  {\doibase 10.1103/PhysRevLett.126.181301} {\bibfield  {journal} {\bibinfo
  {journal} {Phys. Rev. Lett.}\ }\textbf {\bibinfo {volume} {126}},\ \bibinfo
  {pages} {181301} (\bibinfo {year} {2021})},\ \Eprint
  {http://arxiv.org/abs/2012.08785} {arXiv:2012.08785 [gr-qc]} \BibitemShut
  {NoStop}%
\bibitem [{\citenamefont {Liu}\ \emph {et~al.}(2020)\citenamefont {Liu},
  \citenamefont {Zhu}, \citenamefont {Wu}, \citenamefont {Jusufi},
  \citenamefont {Jamil}, \citenamefont {Azreg-A\"\i{}nou},\ and\ \citenamefont
  {Wang}}]{Liu:2020ola}%
  \BibitemOpen
  \bibfield  {author} {\bibinfo {author} {\bibfnamefont {C.}~\bibnamefont
  {Liu}}, \bibinfo {author} {\bibfnamefont {T.}~\bibnamefont {Zhu}}, \bibinfo
  {author} {\bibfnamefont {Q.}~\bibnamefont {Wu}}, \bibinfo {author}
  {\bibfnamefont {K.}~\bibnamefont {Jusufi}}, \bibinfo {author} {\bibfnamefont
  {M.}~\bibnamefont {Jamil}}, \bibinfo {author} {\bibfnamefont
  {M.}~\bibnamefont {Azreg-A\"\i{}nou}}, \ and\ \bibinfo {author}
  {\bibfnamefont {A.}~\bibnamefont {Wang}},\ }\href {\doibase
  10.1103/PhysRevD.101.084001} {\bibfield  {journal} {\bibinfo  {journal}
  {Phys. Rev. D}\ }\textbf {\bibinfo {volume} {101}},\ \bibinfo {pages}
  {084001} (\bibinfo {year} {2020})},\ \bibinfo {note} {[Erratum: Phys.Rev.D
  103, 089902 (2021)]},\ \Eprint {http://arxiv.org/abs/2003.00477}
  {arXiv:2003.00477 [gr-qc]} \BibitemShut {NoStop}%
\bibitem [{\citenamefont {Ghosh}(2015)}]{Ghosh:2014pba}%
  \BibitemOpen
  \bibfield  {author} {\bibinfo {author} {\bibfnamefont {S.~G.}\ \bibnamefont
  {Ghosh}},\ }\href {\doibase 10.1140/epjc/s10052-015-3740-y} {\bibfield
  {journal} {\bibinfo  {journal} {Eur. Phys. J. C}\ }\textbf {\bibinfo {volume}
  {75}},\ \bibinfo {pages} {532} (\bibinfo {year} {2015})},\ \Eprint
  {http://arxiv.org/abs/1408.5668} {arXiv:1408.5668 [gr-qc]} \BibitemShut
  {NoStop}%
\bibitem [{\citenamefont {Ghosh}(2016)}]{Ghosh:2015ovj}%
  \BibitemOpen
  \bibfield  {author} {\bibinfo {author} {\bibfnamefont {S.~G.}\ \bibnamefont
  {Ghosh}},\ }\href {\doibase 10.1140/epjc/s10052-016-4051-7} {\bibfield
  {journal} {\bibinfo  {journal} {Eur. Phys. J. C}\ }\textbf {\bibinfo {volume}
  {76}},\ \bibinfo {pages} {222} (\bibinfo {year} {2016})},\ \Eprint
  {http://arxiv.org/abs/1512.05476} {arXiv:1512.05476 [gr-qc]} \BibitemShut
  {NoStop}%
\bibitem [{\citenamefont {Kumar}\ \emph {et~al.}(2020)\citenamefont {Kumar},
  \citenamefont {Ghosh},\ and\ \citenamefont {Wang}}]{Kumar:2020hgm}%
  \BibitemOpen
  \bibfield  {author} {\bibinfo {author} {\bibfnamefont {R.}~\bibnamefont
  {Kumar}}, \bibinfo {author} {\bibfnamefont {S.~G.}\ \bibnamefont {Ghosh}}, \
  and\ \bibinfo {author} {\bibfnamefont {A.}~\bibnamefont {Wang}},\ }\href
  {\doibase 10.1103/PhysRevD.101.104001} {\bibfield  {journal} {\bibinfo
  {journal} {Phys. Rev. D}\ }\textbf {\bibinfo {volume} {101}},\ \bibinfo
  {pages} {104001} (\bibinfo {year} {2020})},\ \Eprint
  {http://arxiv.org/abs/2001.00460} {arXiv:2001.00460 [gr-qc]} \BibitemShut
  {NoStop}%
\bibitem [{\citenamefont {Kumar}\ and\ \citenamefont
  {Ghosh}(2020)}]{Kumar:2020owy}%
  \BibitemOpen
  \bibfield  {author} {\bibinfo {author} {\bibfnamefont {R.}~\bibnamefont
  {Kumar}}\ and\ \bibinfo {author} {\bibfnamefont {S.~G.}\ \bibnamefont
  {Ghosh}},\ }\href {\doibase 10.1088/1475-7516/2020/07/053} {\bibfield
  {journal} {\bibinfo  {journal} {JCAP}\ }\textbf {\bibinfo {volume} {07}},\
  \bibinfo {pages} {053} (\bibinfo {year} {2020})},\ \Eprint
  {http://arxiv.org/abs/2003.08927} {arXiv:2003.08927 [gr-qc]} \BibitemShut
  {NoStop}%
\bibitem [{\citenamefont {Kumar}\ \emph
  {et~al.}(2022{\natexlab{a}})\citenamefont {Kumar}, \citenamefont {Islam},\
  and\ \citenamefont {Ghosh}}]{Kumar:2021cyl}%
  \BibitemOpen
  \bibfield  {author} {\bibinfo {author} {\bibfnamefont {J.}~\bibnamefont
  {Kumar}}, \bibinfo {author} {\bibfnamefont {S.~U.}\ \bibnamefont {Islam}}, \
  and\ \bibinfo {author} {\bibfnamefont {S.~G.}\ \bibnamefont {Ghosh}},\ }\href
  {\doibase 10.1140/epjc/s10052-022-10357-2} {\bibfield  {journal} {\bibinfo
  {journal} {Eur. Phys. J. C}\ }\textbf {\bibinfo {volume} {82}},\ \bibinfo
  {pages} {443} (\bibinfo {year} {2022}{\natexlab{a}})},\ \Eprint
  {http://arxiv.org/abs/2109.04450} {arXiv:2109.04450 [gr-qc]} \BibitemShut
  {NoStop}%
\bibitem [{\citenamefont {Islam}\ and\ \citenamefont
  {Ghosh}(2021)}]{Islam:2021dyk}%
  \BibitemOpen
  \bibfield  {author} {\bibinfo {author} {\bibfnamefont {S.~U.}\ \bibnamefont
  {Islam}}\ and\ \bibinfo {author} {\bibfnamefont {S.~G.}\ \bibnamefont
  {Ghosh}},\ }\href {\doibase 10.1103/PhysRevD.103.124052} {\bibfield
  {journal} {\bibinfo  {journal} {Phys. Rev. D}\ }\textbf {\bibinfo {volume}
  {103}},\ \bibinfo {pages} {124052} (\bibinfo {year} {2021})},\ \Eprint
  {http://arxiv.org/abs/2102.08289} {arXiv:2102.08289 [gr-qc]} \BibitemShut
  {NoStop}%
\bibitem [{\citenamefont {Hawking}\ and\ \citenamefont
  {Ellis}(2011)}]{Hawking:1973uf}%
  \BibitemOpen
  \bibfield  {author} {\bibinfo {author} {\bibfnamefont {S.~W.}\ \bibnamefont
  {Hawking}}\ and\ \bibinfo {author} {\bibfnamefont {G.~F.~R.}\ \bibnamefont
  {Ellis}},\ }\href {\doibase 10.1017/CBO9780511524646} {\emph {\bibinfo
  {title} {{The Large Scale Structure of Space-Time}}}},\ Cambridge Monographs
  on Mathematical Physics\ (\bibinfo  {publisher} {Cambridge University
  Press},\ \bibinfo {year} {2011})\BibitemShut {NoStop}%
\bibitem [{\citenamefont {Ghosh}\ and\ \citenamefont
  {Kothawala}(2008)}]{Ghosh:2008zza}%
  \BibitemOpen
  \bibfield  {author} {\bibinfo {author} {\bibfnamefont {S.~G.}\ \bibnamefont
  {Ghosh}}\ and\ \bibinfo {author} {\bibfnamefont {D.}~\bibnamefont
  {Kothawala}},\ }\href {\doibase 10.1007/s10714-007-0511-6} {\bibfield
  {journal} {\bibinfo  {journal} {Gen. Rel. Grav.}\ }\textbf {\bibinfo {volume}
  {40}},\ \bibinfo {pages} {9--21} (\bibinfo {year} {2008})},\ \Eprint
  {http://arxiv.org/abs/0801.4342} {arXiv:0801.4342 [gr-qc]} \BibitemShut
  {NoStop}%
\bibitem [{\citenamefont {Kothawala}\ and\ \citenamefont
  {Ghosh}(2004)}]{Kothawala:2004fy}%
  \BibitemOpen
  \bibfield  {author} {\bibinfo {author} {\bibfnamefont {D.}~\bibnamefont
  {Kothawala}}\ and\ \bibinfo {author} {\bibfnamefont {S.~G.}\ \bibnamefont
  {Ghosh}},\ }\href {\doibase 10.1103/PhysRevD.70.104010} {\bibfield  {journal}
  {\bibinfo  {journal} {Phys. Rev. D}\ }\textbf {\bibinfo {volume} {70}},\
  \bibinfo {pages} {104010} (\bibinfo {year} {2004})},\ \Eprint
  {http://arxiv.org/abs/1007.2500} {arXiv:1007.2500 [gr-qc]} \BibitemShut
  {NoStop}%
\bibitem [{\citenamefont
  {Azreg-A\"\i{}nou}(2014{\natexlab{a}})}]{Azreg-Ainou:2014pra}%
  \BibitemOpen
  \bibfield  {author} {\bibinfo {author} {\bibfnamefont {M.}~\bibnamefont
  {Azreg-A\"\i{}nou}},\ }\href {\doibase 10.1103/PhysRevD.90.064041} {\bibfield
   {journal} {\bibinfo  {journal} {Phys. Rev. D}\ }\textbf {\bibinfo {volume}
  {90}},\ \bibinfo {pages} {064041} (\bibinfo {year} {2014}{\natexlab{a}})},\
  \Eprint {http://arxiv.org/abs/1405.2569} {arXiv:1405.2569 [gr-qc]}
  \BibitemShut {NoStop}%
\bibitem [{\citenamefont
  {Azreg-A\"\i{}nou}(2014{\natexlab{b}})}]{Azreg-Ainou:2014aqa}%
  \BibitemOpen
  \bibfield  {author} {\bibinfo {author} {\bibfnamefont {M.}~\bibnamefont
  {Azreg-A\"\i{}nou}},\ }\href {\doibase 10.1140/epjc/s10052-014-2865-8}
  {\bibfield  {journal} {\bibinfo  {journal} {Eur. Phys. J. C}\ }\textbf
  {\bibinfo {volume} {74}},\ \bibinfo {pages} {2865} (\bibinfo {year}
  {2014}{\natexlab{b}})},\ \Eprint {http://arxiv.org/abs/1401.4292}
  {arXiv:1401.4292 [gr-qc]} \BibitemShut {NoStop}%
\bibitem [{\citenamefont {Newman}\ and\ \citenamefont
  {Janis}(1965)}]{Newman:1965tw}%
  \BibitemOpen
  \bibfield  {author} {\bibinfo {author} {\bibfnamefont {E.~T.}\ \bibnamefont
  {Newman}}\ and\ \bibinfo {author} {\bibfnamefont {A.~I.}\ \bibnamefont
  {Janis}},\ }\href {\doibase 10.1063/1.1704350} {\bibfield  {journal}
  {\bibinfo  {journal} {J. Math. Phys.}\ }\textbf {\bibinfo {volume} {6}},\
  \bibinfo {pages} {915--917} (\bibinfo {year} {1965})}\BibitemShut {NoStop}%
\bibitem [{\citenamefont {Johannsen}\ and\ \citenamefont
  {Psaltis}(2011)}]{Johannsen:2011dh}%
  \BibitemOpen
  \bibfield  {author} {\bibinfo {author} {\bibfnamefont {T.}~\bibnamefont
  {Johannsen}}\ and\ \bibinfo {author} {\bibfnamefont {D.}~\bibnamefont
  {Psaltis}},\ }\href {\doibase 10.1103/PhysRevD.83.124015} {\bibfield
  {journal} {\bibinfo  {journal} {Phys. Rev. D}\ }\textbf {\bibinfo {volume}
  {83}},\ \bibinfo {pages} {124015} (\bibinfo {year} {2011})},\ \Eprint
  {http://arxiv.org/abs/1105.3191} {arXiv:1105.3191 [gr-qc]} \BibitemShut
  {NoStop}%
\bibitem [{\citenamefont {Jusufi}\ \emph {et~al.}(2020)\citenamefont {Jusufi},
  \citenamefont {Jamil}, \citenamefont {Chakrabarty}, \citenamefont {Wu},
  \citenamefont {Bambi},\ and\ \citenamefont {Wang}}]{Jusufi:2019caq}%
  \BibitemOpen
  \bibfield  {author} {\bibinfo {author} {\bibfnamefont {K.}~\bibnamefont
  {Jusufi}}, \bibinfo {author} {\bibfnamefont {M.}~\bibnamefont {Jamil}},
  \bibinfo {author} {\bibfnamefont {H.}~\bibnamefont {Chakrabarty}}, \bibinfo
  {author} {\bibfnamefont {Q.}~\bibnamefont {Wu}}, \bibinfo {author}
  {\bibfnamefont {C.}~\bibnamefont {Bambi}}, \ and\ \bibinfo {author}
  {\bibfnamefont {A.}~\bibnamefont {Wang}},\ }\href {\doibase
  10.1103/PhysRevD.101.044035} {\bibfield  {journal} {\bibinfo  {journal}
  {Phys. Rev. D}\ }\textbf {\bibinfo {volume} {101}},\ \bibinfo {pages}
  {044035} (\bibinfo {year} {2020})},\ \Eprint
  {http://arxiv.org/abs/1911.07520} {arXiv:1911.07520 [gr-qc]} \BibitemShut
  {NoStop}%
\bibitem [{\citenamefont {Ghosh}\ and\ \citenamefont
  {Maharaj}(2015)}]{Ghosh:2014hea}%
  \BibitemOpen
  \bibfield  {author} {\bibinfo {author} {\bibfnamefont {S.~G.}\ \bibnamefont
  {Ghosh}}\ and\ \bibinfo {author} {\bibfnamefont {S.~D.}\ \bibnamefont
  {Maharaj}},\ }\href {\doibase 10.1140/epjc/s10052-014-3222-7} {\bibfield
  {journal} {\bibinfo  {journal} {Eur. Phys. J. C}\ }\textbf {\bibinfo {volume}
  {75}},\ \bibinfo {pages} {7} (\bibinfo {year} {2015})},\ \Eprint
  {http://arxiv.org/abs/1410.4043} {arXiv:1410.4043 [gr-qc]} \BibitemShut
  {NoStop}%
\bibitem [{\citenamefont {Moffat}(2015)}]{Moffat:2014aja}%
  \BibitemOpen
  \bibfield  {author} {\bibinfo {author} {\bibfnamefont {J.~W.}\ \bibnamefont
  {Moffat}},\ }\href {\doibase 10.1140/epjc/s10052-015-3405-x} {\bibfield
  {journal} {\bibinfo  {journal} {Eur. Phys. J. C}\ }\textbf {\bibinfo {volume}
  {75}},\ \bibinfo {pages} {175} (\bibinfo {year} {2015})},\ \Eprint
  {http://arxiv.org/abs/1412.5424} {arXiv:1412.5424 [gr-qc]} \BibitemShut
  {NoStop}%
\bibitem [{\citenamefont {Hansen}\ and\ \citenamefont
  {Yunes}(2013)}]{Hansen:2013owa}%
  \BibitemOpen
  \bibfield  {author} {\bibinfo {author} {\bibfnamefont {D.}~\bibnamefont
  {Hansen}}\ and\ \bibinfo {author} {\bibfnamefont {N.}~\bibnamefont {Yunes}},\
  }\href {\doibase 10.1103/PhysRevD.88.104020} {\bibfield  {journal} {\bibinfo
  {journal} {Phys. Rev. D}\ }\textbf {\bibinfo {volume} {88}},\ \bibinfo
  {pages} {104020} (\bibinfo {year} {2013})},\ \Eprint
  {http://arxiv.org/abs/1308.6631} {arXiv:1308.6631 [gr-qc]} \BibitemShut
  {NoStop}%
\bibitem [{\citenamefont {Kumar}\ \emph
  {et~al.}(2022{\natexlab{b}})\citenamefont {Kumar}, \citenamefont {Islam},\
  and\ \citenamefont {Ghosh}}]{Kumar:2022fqo}%
  \BibitemOpen
  \bibfield  {author} {\bibinfo {author} {\bibfnamefont {J.}~\bibnamefont
  {Kumar}}, \bibinfo {author} {\bibfnamefont {S.~U.}\ \bibnamefont {Islam}}, \
  and\ \bibinfo {author} {\bibfnamefont {S.~G.}\ \bibnamefont {Ghosh}},\
  }\href@noop {} {\  (\bibinfo {year} {2022}{\natexlab{b}})},\ \Eprint
  {http://arxiv.org/abs/2209.04240} {arXiv:2209.04240 [gr-qc]} \BibitemShut
  {NoStop}%
\bibitem [{\citenamefont {Afrin}\ and\ \citenamefont
  {Ghosh}(2022)}]{Afrin:2021wlj}%
  \BibitemOpen
  \bibfield  {author} {\bibinfo {author} {\bibfnamefont {M.}~\bibnamefont
  {Afrin}}\ and\ \bibinfo {author} {\bibfnamefont {S.~G.}\ \bibnamefont
  {Ghosh}},\ }\href {\doibase 10.3847/1538-4357/ac6dda} {\bibfield  {journal}
  {\bibinfo  {journal} {Astrophys. J.}\ }\textbf {\bibinfo {volume} {932}},\
  \bibinfo {pages} {51} (\bibinfo {year} {2022})},\ \Eprint
  {http://arxiv.org/abs/2110.05258} {arXiv:2110.05258 [gr-qc]} \BibitemShut
  {NoStop}%
\bibitem [{\citenamefont {Ghosh}\ and\ \citenamefont
  {Walia}(2022)}]{Ghosh:2022jfi}%
  \BibitemOpen
  \bibfield  {author} {\bibinfo {author} {\bibfnamefont {S.~G.}\ \bibnamefont
  {Ghosh}}\ and\ \bibinfo {author} {\bibfnamefont {R.~K.}\ \bibnamefont
  {Walia}},\ }in\ \href@noop {} {\emph {\bibinfo {booktitle} {{16th Marcel
  Grossmann Meeting on~Recent Developments in Theoretical and Experimental
  General Relativity, Astrophysics and Relativistic Field Theories}}}}\
  (\bibinfo {year} {2022})\ \Eprint {http://arxiv.org/abs/2203.07775}
  {arXiv:2203.07775 [gr-qc]} \BibitemShut {NoStop}%
\bibitem [{\citenamefont {Walia}\ \emph {et~al.}(2022)\citenamefont {Walia},
  \citenamefont {Maharaj},\ and\ \citenamefont {Ghosh}}]{Walia:2021emv}%
  \BibitemOpen
  \bibfield  {author} {\bibinfo {author} {\bibfnamefont {R.~K.}\ \bibnamefont
  {Walia}}, \bibinfo {author} {\bibfnamefont {S.~D.}\ \bibnamefont {Maharaj}},
  \ and\ \bibinfo {author} {\bibfnamefont {S.~G.}\ \bibnamefont {Ghosh}},\
  }\href {\doibase 10.1140/epjc/s10052-022-10451-5} {\bibfield  {journal}
  {\bibinfo  {journal} {Eur. Phys. J. C}\ }\textbf {\bibinfo {volume} {82}},\
  \bibinfo {pages} {547} (\bibinfo {year} {2022})},\ \Eprint
  {http://arxiv.org/abs/2109.08055} {arXiv:2109.08055 [gr-qc]} \BibitemShut
  {NoStop}%
\bibitem [{\citenamefont {Akiyama}\ \emph
  {et~al.}(2019{\natexlab{a}})\citenamefont {Akiyama} \emph
  {et~al.}}]{EventHorizonTelescope:2019dse}%
  \BibitemOpen
  \bibfield  {author} {\bibinfo {author} {\bibfnamefont {K.}~\bibnamefont
  {Akiyama}} \emph {et~al.} (\bibinfo {collaboration} {Event Horizon
  Telescope}),\ }\href {\doibase 10.3847/2041-8213/ab0ec7} {\bibfield
  {journal} {\bibinfo  {journal} {Astrophys. J. Lett.}\ }\textbf {\bibinfo
  {volume} {875}},\ \bibinfo {pages} {L1} (\bibinfo {year}
  {2019}{\natexlab{a}})},\ \Eprint {http://arxiv.org/abs/1906.11238}
  {arXiv:1906.11238 [astro-ph.GA]} \BibitemShut {NoStop}%
\bibitem [{\citenamefont {Akiyama}\ \emph
  {et~al.}(2019{\natexlab{b}})\citenamefont {Akiyama} \emph
  {et~al.}}]{EventHorizonTelescope:2019ggy}%
  \BibitemOpen
  \bibfield  {author} {\bibinfo {author} {\bibfnamefont {K.}~\bibnamefont
  {Akiyama}} \emph {et~al.} (\bibinfo {collaboration} {Event Horizon
  Telescope}),\ }\href {\doibase 10.3847/2041-8213/ab1141} {\bibfield
  {journal} {\bibinfo  {journal} {Astrophys. J. Lett.}\ }\textbf {\bibinfo
  {volume} {875}},\ \bibinfo {pages} {L6} (\bibinfo {year}
  {2019}{\natexlab{b}})},\ \Eprint {http://arxiv.org/abs/1906.11243}
  {arXiv:1906.11243 [astro-ph.GA]} \BibitemShut {NoStop}%
\bibitem [{\citenamefont {Akiyama}\ \emph
  {et~al.}(2022{\natexlab{a}})\citenamefont {Akiyama} \emph
  {et~al.}}]{EventHorizonTelescope:2022xnr}%
  \BibitemOpen
  \bibfield  {author} {\bibinfo {author} {\bibfnamefont {K.}~\bibnamefont
  {Akiyama}} \emph {et~al.} (\bibinfo {collaboration} {Event Horizon
  Telescope}),\ }\href {\doibase 10.3847/2041-8213/ac6674} {\bibfield
  {journal} {\bibinfo  {journal} {Astrophys. J. Lett.}\ }\textbf {\bibinfo
  {volume} {930}},\ \bibinfo {pages} {L12} (\bibinfo {year}
  {2022}{\natexlab{a}})}\BibitemShut {NoStop}%
\bibitem [{\citenamefont {Akiyama}\ \emph
  {et~al.}(2022{\natexlab{b}})\citenamefont {Akiyama} \emph
  {et~al.}}]{EventHorizonTelescope:2022xqj}%
  \BibitemOpen
  \bibfield  {author} {\bibinfo {author} {\bibfnamefont {K.}~\bibnamefont
  {Akiyama}} \emph {et~al.} (\bibinfo {collaboration} {Event Horizon
  Telescope}),\ }\href {\doibase 10.3847/2041-8213/ac6756} {\bibfield
  {journal} {\bibinfo  {journal} {Astrophys. J. Lett.}\ }\textbf {\bibinfo
  {volume} {930}},\ \bibinfo {pages} {L17} (\bibinfo {year}
  {2022}{\natexlab{b}})}\BibitemShut {NoStop}%
\bibitem [{\citenamefont {Islam}\ \emph {et~al.}(2022)\citenamefont {Islam},
  \citenamefont {Kumar}, \citenamefont {Kumar},\ and\ \citenamefont
  {Ghosh}}]{Shafqat:2022}%
  \BibitemOpen
  \bibfield  {author} {\bibinfo {author} {\bibfnamefont {S.~U.}\ \bibnamefont
  {Islam}}, \bibinfo {author} {\bibfnamefont {J.}~\bibnamefont {Kumar}},
  \bibinfo {author} {\bibfnamefont {R.}~\bibnamefont {Kumar}}, \ and\ \bibinfo
  {author} {\bibfnamefont {S.~G.}\ \bibnamefont {Ghosh}},\ }\href@noop {}
  {\enquote {\bibinfo {title} {{Investigating Loop Quantum Gravity with EHT
  Observational Effects of Rotating Black holes}},}\ } (\bibinfo {year}
  {2022}),\ \bibinfo {note} {preprint}\BibitemShut {NoStop}%
\end{thebibliography}%
\bibliographystyle{apsrev4-1}
\end{document}